\documentclass[aps,prl,nobalancelastpage,twocolumn,superscriptaddress,reprint]{revtex4-2}
\usepackage{ulem}
\usepackage{hyperref}
\hypersetup{
    colorlinks=true,
    linkcolor=blue,
    filecolor=gray,      
    urlcolor=blue,
    citecolor=blue,
}
\usepackage{graphicx}
\usepackage{tabularx}
\usepackage{booktabs}
\usepackage{bm, physics}

\begin{document}

\title{Deep-Learning Database of Density Functional Theory Hamiltonians for Twisted Materials}
\affiliation{State Key Laboratory of Low Dimensional Quantum Physics and Department of Physics, Tsinghua University, Beijing, 100084, China}
\affiliation{Institute for Advanced Study, Tsinghua University, Beijing 100084, China}
\affiliation{School of Physics, Peking University, Beijing 100871, China}
\affiliation{Frontier Science Center for Quantum Information, Beijing, China}
\affiliation{Beijing Academy of Quantum Information Sciences, Beijing 100193, China}
\affiliation{RIKEN Center for Emergent Matter Science (CEMS), Wako, Saitama 351-0198, Japan}
\affiliation{These authors contributed equally}

\author{Ting \surname{Bao}}
\affiliation{State Key Laboratory of Low Dimensional Quantum Physics and Department of Physics, Tsinghua University, Beijing, 100084, China}
\affiliation{These authors contributed equally}

\author{Runzhang \surname{Xu}}
\email{xurz@mail.tsinghua.edu.cn}
\affiliation{State Key Laboratory of Low Dimensional Quantum Physics and Department of Physics, Tsinghua University, Beijing, 100084, China}
\affiliation{These authors contributed equally}

\author{He \surname{Li}}
\affiliation{State Key Laboratory of Low Dimensional Quantum Physics and Department of Physics, Tsinghua University, Beijing, 100084, China}
\affiliation{Institute for Advanced Study, Tsinghua University, Beijing 100084, China}

\author{Xiaoxun \surname{Gong}}
\affiliation{School of Physics, Peking University, Beijing 100871, China}

\author{Zechen \surname{Tang}}
\affiliation{State Key Laboratory of Low Dimensional Quantum Physics and Department of Physics, Tsinghua University, Beijing, 100084, China}

\author{Jingheng \surname{Fu}}
\affiliation{State Key Laboratory of Low Dimensional Quantum Physics and Department of Physics, Tsinghua University, Beijing, 100084, China}

\author{Wenhui \surname{Duan}}
\email{duanw@tsinghua.edu.cn}
\affiliation{State Key Laboratory of Low Dimensional Quantum Physics and Department of Physics, Tsinghua University, Beijing, 100084, China}
\affiliation{Institute for Advanced Study, Tsinghua University, Beijing 100084, China}
\affiliation{Frontier Science Center for Quantum Information, Beijing, China}
\affiliation{Beijing Academy of Quantum Information Sciences, Beijing 100193, China}

\author{Yong \surname{Xu}}
\email{yongxu@mail.tsinghua.edu.cn}
\affiliation{State Key Laboratory of Low Dimensional Quantum Physics and Department of Physics, Tsinghua University, Beijing, 100084, China}
\affiliation{Frontier Science Center for Quantum Information, Beijing, China}
\affiliation{RIKEN Center for Emergent Matter Science (CEMS), Wako, Saitama 351-0198, Japan}

\begin{abstract}
Moiré-twisted materials have garnered significant research interest due to their distinctive properties and intriguing physics. However, conducting first-principles studies on such materials faces challenges, notably the formidable computational cost associated with simulating ultra-large twisted structures. This obstacle impedes the construction of a twisted materials database crucial for data-driven materials discovery. Here, by using high-throughput calculations and state-of-the-art neural network methods, we construct a Deep-learning Database of density functional theory (DFT) Hamiltonians for Twisted materials named DDHT. The DDHT database comprises trained neural-network models of over a hundred homo-bilayer and hetero-bilayer moiré-twisted materials. These models enable accurate prediction of the DFT Hamiltonian for these materials across arbitrary twist angles,  with an averaged mean absolute error of approximately 1.0 meV or lower. The database facilitates the exploration of flat bands and correlated materials platforms within ultra-large twisted structures.
\end{abstract}

\maketitle
\section{Introduction}
\begin{figure*}[!th]
    \centering
    \includegraphics[width=1.00\textwidth]{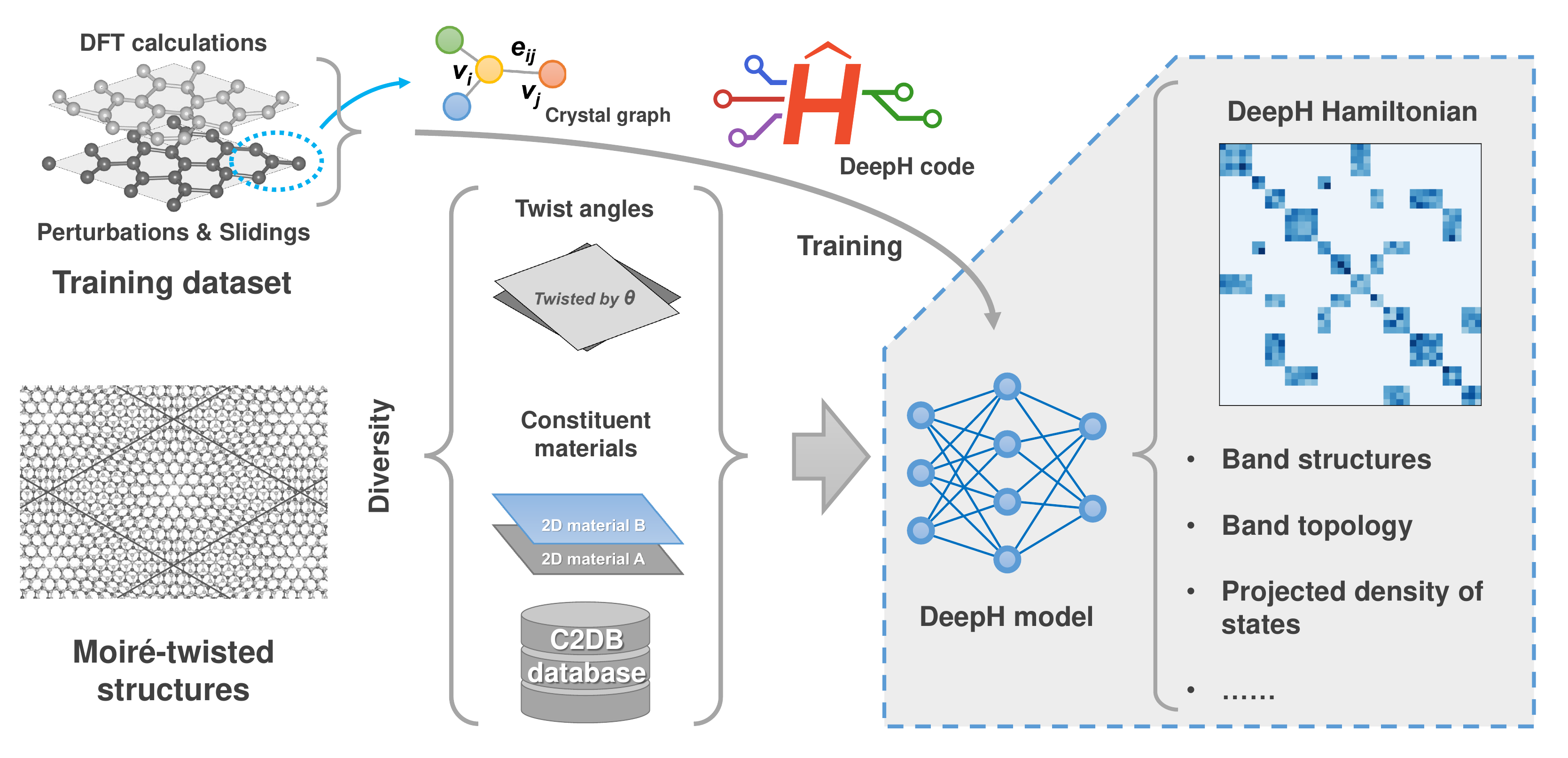}
    \caption{Schematic diagram of construction and application of DDHT database. Uniformly distributed interlayer slidings together with random atomic perturbations are introduced in the non-twisted bilayer supercells to construct training datasets. Graph neural networks are employed here, using materials structures as input and their DFT Hamiltonians as output. Deep neural network models are trained by DFT calculation results of training structures, and then generalized to study diverse moiré-twisted structures with varying twist angles and constituent materials selected from C2DB. From the DFT Hamiltonians predicted by DeepH, many physical properties can be derived, including band structure, band topology, projected density of states, etc.}
    \label{fig:diagram} 
\end{figure*}



The discovery of correlated states in magic-angle twisted bilayer graphene (TBG) \cite{Cao2018_1,Cao2018_2} has sparked great research interest in the emergent physics and exotic properties in moiré-twisted two-dimensional (2D) materials. In moiré twist systems, the relative twist rotation between the two constituent van der Waals (vdW) layers forms moiré patterns with tunable periodicity, generating a moiré potential field that strongly influences the electronic structures and introduces exotic quantum effects such as correlated states, band topology, spin textures, and ferroelectric polarization \cite{sun2024twisted}. These twist-induced interesting underlying physics have been intensively studied on TBG, twisted bilayer CrI$_3$\cite{Song2021,Xu2021}, twisted bilayer BN (TBBN)\cite{ViznerStern2021,Yasuda2021}, twisted transition metal dichalcogenides (TTMDs) \cite{devakul2021magic} and so on.
Other twist-induced properties, like moiré excitons \cite{Jin2019,Tran2019,Alexeev2019,Shimazaki2020}, Wigner crystal states \cite{Regan2020}, and spin- and bond-density waves \cite{Xian2019,Kennes2020} further expand the knowledge of twisted materials. 
These intriguing effects and  physics, coupled with the rich external modulations, position twisted materials not only as ideal quantum simulators \cite{Kennes2021} in condensed-matter research but also as practical platforms for twistronic applications. 

Despite the growing interest in twisted materials and the related physics, the field faces fundamental challenges from both experimental and theoretical perspectives~\cite{Carr2020}. In experiments, the synthesis of high-quality twisted samples and the precise control of twist angles are rather challenging. From theoretical and calculation perspective, various kinds of numerical and theoretical methods, such as DFT, tight-binding method, and continuum model, are employed to make predictions for guiding experiments. However, for twisted materials, all these theoretical methods encounter the vital efficiency-accuracy dilemma. For instance, DFT methods scales as $\mathcal{O}(N^3)$ with $N$ being number of atoms per periodic unit cell, which limits the sizes of twisted structures to be studied. The empirical tight-binding methods or continuum models are much more efficient, but suffers from the critical problem of low accuracy. The revolutionary success of artificial intelligence, especially deep learning neural networks, in many fields \cite{Senior2020,Zhang2018,behler2007generalized} brings solution to this dilemma. In previous work DeepH \cite{li_deep-learning_2022}, Li et.al. show that a neural network (NN) model trained on small structures can be generalized to tackle the DFT Hamiltonians of large twisted structures at significantly reduced computational cost. However, only TBG, twisted bilayer bismuthene and twisted MoS$_2$ are studied using DeepH, which is not enough to validate the effectiveness. A more systematic learning on DeepH methods is in urgent need, which is not only to check the validness and universality of DeepH but also to meet the demand of getting the DFT-level Hamiltonians for more materials and boost further research.

Here, we report the construction of a deep-learning database of DFT Hamiltonians for twisted materials named DDHT as shown in Fig. \ref{fig:diagram}. DDHT currently covers DeepH NN models for 124 types of 2D homo-bilayer and 5 types of hetero-bilayer twist materials and can efficiently predict accurate DFT Hamiltonians of their twisted structures across arbitrary twist angles. For each type of material, the NN model is trained by the state-of-the-art deep-learning representation of DFT Hamiltonians implemented in the E3-equivariant DeepH method \cite{gong2023general,li_deep-learning_2022}. We demonstrate the high reliability and DFT-comparable accuracy of all NN models in our database by the very low validation loss achieved in the NN training process and using the DFT-calculated Hamiltonians at selected large twist angles as benchmark. Furthermore, taking 2D black phosphorene (bP) as an exhibition, we use the corresponding NN model in DDHT to obtain electronic band structure and find the ultra-flat valence bands in twisted bilayer bP down to twist angles inaccessed by conventional DFT methods, showing the effectiveness and efficiency of our database in getting exotic electronic properties of ultra-large twisted materials. In our exhibition of twisted bP and probably many other twisted materials systems, the flat bands usually have bandwidth droping to only a few meV at small-angle limit ($<$2.0 $^\circ$) and their real space projection can be highly localized, giving rise to rich exotic properties from doping modulation and strong correlation. DDHT not only provides an optimal parameter strategy for NN training and strongly supports the universality of DeepH method, but also offers prior DFT-level knowledge of Hamiltonians and electronic structures for researchers in the field of twisted materials and twistronics and thus greatly accelerate the exploration in this field.

It is worth mentioning that unlike traditional material databases like Materials Project \cite{Jain2013}, ICSD \cite{hellenbrandt2004inorganic} or C2DB \cite{haastrupComputational2018,gjerdingRecent2021}, which includes specific materials corresponding to discrete points in the materials parameter space, our DDHT is applicable to twisted structures at arbitrary twist angles and interlayer fluctuation of provided types of material, covering wide areas in the materials parameter space with infinite single structure cases inside. Furthermore, considering each model in the DDHT as an expert and adding a simple classifier as the gate on types of material, out DDHT itself is actually serving as a mixture of experts (MoE) model \cite{jacobs1991adaptive} with each NN item experienced in one type of twisted material. Thus DDHT has dualism of database and MoE model, we suppose this dualism is a new formunism and will be a necessity of future databases in the AI era.

\section{Results}
\subsection{Construction of Database}


\begin{figure*}
    \centering
    \includegraphics[width= 1.00\textwidth]{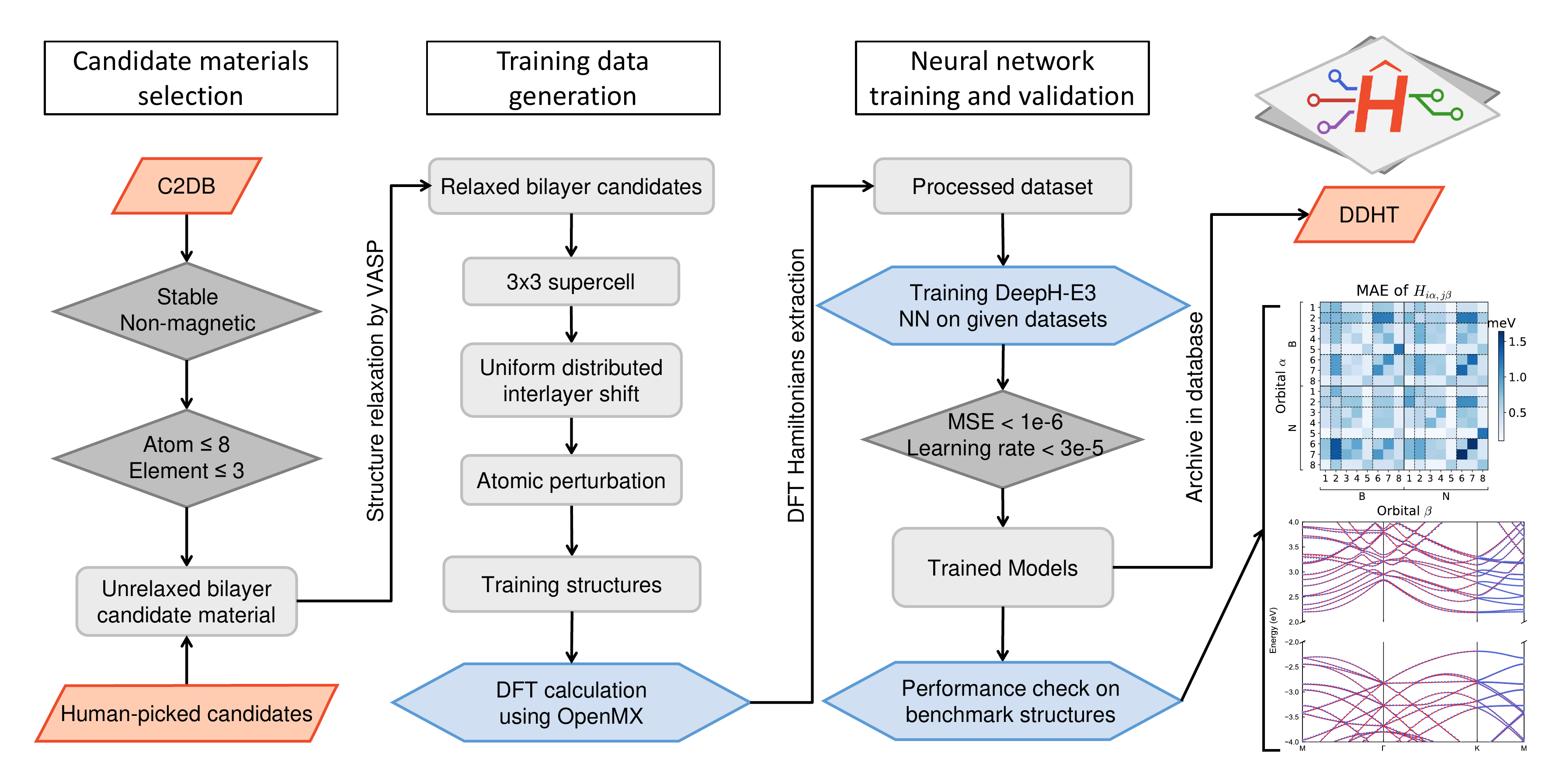}
    \caption{Workflow of constructing deep-learning database. All the steps are classified into three parts: candidate selection, dataset generation, and neural network training and validation. The parameters used in construction are explicitly displayed by typical values for key steps. The eventually obtained DDHT database can be further used for unlimited cases.}
    \label{fig:workflow}
\end{figure*}

Candidate 2D materials considered in the current database are adopted from C2DB \cite{haastrupComputational2018,gjerdingRecent2021}. They are further screened to construct twisted materials structures and training datasets (left part of Fig. \ref{fig:workflow}). In total, 124 types of 2D materials (Table S1) in the C2DB are selected following three criteria: non-magnetic, thermally and dynamically stable, structurally simple ($\le$ 8 atoms per unit cell, $\le$ 3 elements). Their bilayer unit cells are constructed and fully relaxed by DFT calculations using Vienna ab initio simulation package (VASP) \cite{kresse1996_1,kresse1996efficient} (see Note 1 in Supporting Information for details of screening and relaxation). This allows us to obtain the optimal structural parameters, including lattice constants, stacking orders, and interlayer spacings. These structural parameters will be used to construct the non-twisted supercell structures in the datasets and the corresponding 124 (5) types of to-be-predicted moiré-twisted homo-bilayer (hetero-bilayer) structures with varying twist angles $\theta$s. These candidates not only cover most of the intensively studied 2D materials and their twisted structures, such as graphene, black phosphorene (bP), and transition metal dichalcogenides (TMDCs), but also include most light elements in the periodic table (Fig. S1).

The schematic diagram of constructing and using our database is demonstrated in Fig. \ref{fig:diagram} using TBG as an example material. Here, the construction involves both the preparation of appropriate datasets and the training of deep neural network models. The usage refers to the prediction of DFT Hamiltonians for twisted materials with any given twist angle as well as various kinds of derived physical properties. We first discuss the preparation of training datasets. Moiré-twisted materials host rich structural diversity in terms of twist angle and constituent materials, which should be captured by atomic structures in the training datasets for accurate ${\mathcal{R}} \mapsto H_{\text{DeepH-E3}}$ mapping by the trained models. This requirement is satisfied by preparing atomic structures in datasets respectively (or separately) for each type of twisted materials, and introducing diverse enough variations to local atomic positions and interlayer stackings to cover all possible local environments in twisted bilayer structures. For instance, TBG and TBBN have different constituent 2D materials (graphene and BN), and the non-twisted supercells in their datasets are generated using graphene and BN bilayer unit cells respectively, while the twisted heterobilayer graphene/BN have heterogeneity in layers and thus uses graphene/BN unit cells for generating the dataset. The relative rotations induced by twisting between the two vdW layers can be treated as locally approximated interlayer shear slidings combined with perturbations to atomic positions in non-twisted bilayer structures. Therefore, for each type of twisted material, the atomic structures in its dataset are prepared ("Training data generation" part in Fig. \ref{fig:workflow}) by 1) making non-twisted supercells with sufficient large lateral size from the most optimal bilayer unit cells to include all neighbors within cut-off radius $R_c$ of local atomic basis and also to avoid interactions from periodic images, and 2) introducing uniform distributed interlayer slidings between the two vdW layers and random perturbations to the atom positions in these non-twisted supercells to cover the local structural diversity in twisted structures, with the mesh density of slidings and the upper limit of perturbations determined based on balanced accuracy and efficiency. The local-atomic-basis Hamiltonians $H_\text{DFT}$ of all these non-twisted, perturbed, and shifted supercell structures in the dataset are calculated self-consistently by first-principles DFT methods and further processed to extract the decomposed hopping matrix segments $\mathcal{H}_{ij}$ that are in one-to-one correspondence with local atomic structures $\mathcal{R}_{ij}$ within $R_c$. After representing $\mathcal{R}_{ij}$ by crystal graphs \cite{PhysRevLett.120.145301}, we finally prepared the complete training dataset for each twisted material by incorporating these $\mathcal{H}_{ij}$--$\mathcal{R}_{ij}$ pairs. 

Based on each training dataset, an E(3)-equivariant deep neural network model is built and trained ("Neural network
training and validation" part in Fig. \ref{fig:workflow}) using the state-of-the-art DeepH-E3 code \cite{gong2023general,li_deep-learning_2022}. To ensure the high accuracy and generalizability of the model, the training process runs for a sufficient number of epochs and stops when the validation loss reaches a strict criterion (see Sec. METHODS for details). Based on a self-made high-throughput computation framework, the dataset preparation and the model training for all $124+5$ types of twisted materials are carried out in high efficiency, which in the end output $124+5$ E(3)-equivariant deep neural network models and their corresponding non-twisted bilayer unit cells that both constitute the database. The DFT calculations in preparation of training datasets and the process of model training are both, to some extend, computationally expansive. Typically, for each type of material, they will take 1-3 days on a 64-core CPU node and 2-7 days on a NVIDIA V100 GPU card. However, the resulting well-trained DeepH-E3 models in our database have strong generalizability and universality over a wide range of $\theta$s (from 21.79$^\circ$ all the way down to less than 1$^\circ$) and are capable of predicting DFT-accuracy $H_\text{DeepH-E3}$ of ultra-large (ultra small-angle) twisted structures beyond the reach of common DFT codes with an acceptable cost. For example, the computational time of predicting $H_\text{DeepH-E3}$ scales only linearly with the number of atoms in lattice cells, allowing for the calculation of DFT-accuracy Hamiltonian for magic-angle TBG in just about 100 seconds \cite{li_deep-learning_2022}. 

Other than the direct usage of our database in predicting $H_\text{DeepH-E3}$ by direct neural-network mapping for twisted structures with arbitrary $\theta$s and situation of interlayer fluctuation, several important physical properties can be further calculated from the predicted $H_\text{DeepH-E3}$ sparse matrices (dashed box on the right in Fig. \ref{fig:diagram}), which may include the eigen values and states, band structures, total and projected density of states (DOS), and possible correlated flat bands near the Fermi level and their real-space projections. It should be emphasized that our database, serving as a MoE for DFT Hamiltonian prediction, offers an ultra-efficient way (enhanced by orders of magnitude \cite{li_deep-learning_2022}) to bypass the most efficiency-limiting DFT calculations in computational research flow in twisted materials and still output Hamiltonians with similar accuracy.

\begin{figure*}
    \centering
    \includegraphics[width=1.00\textwidth]{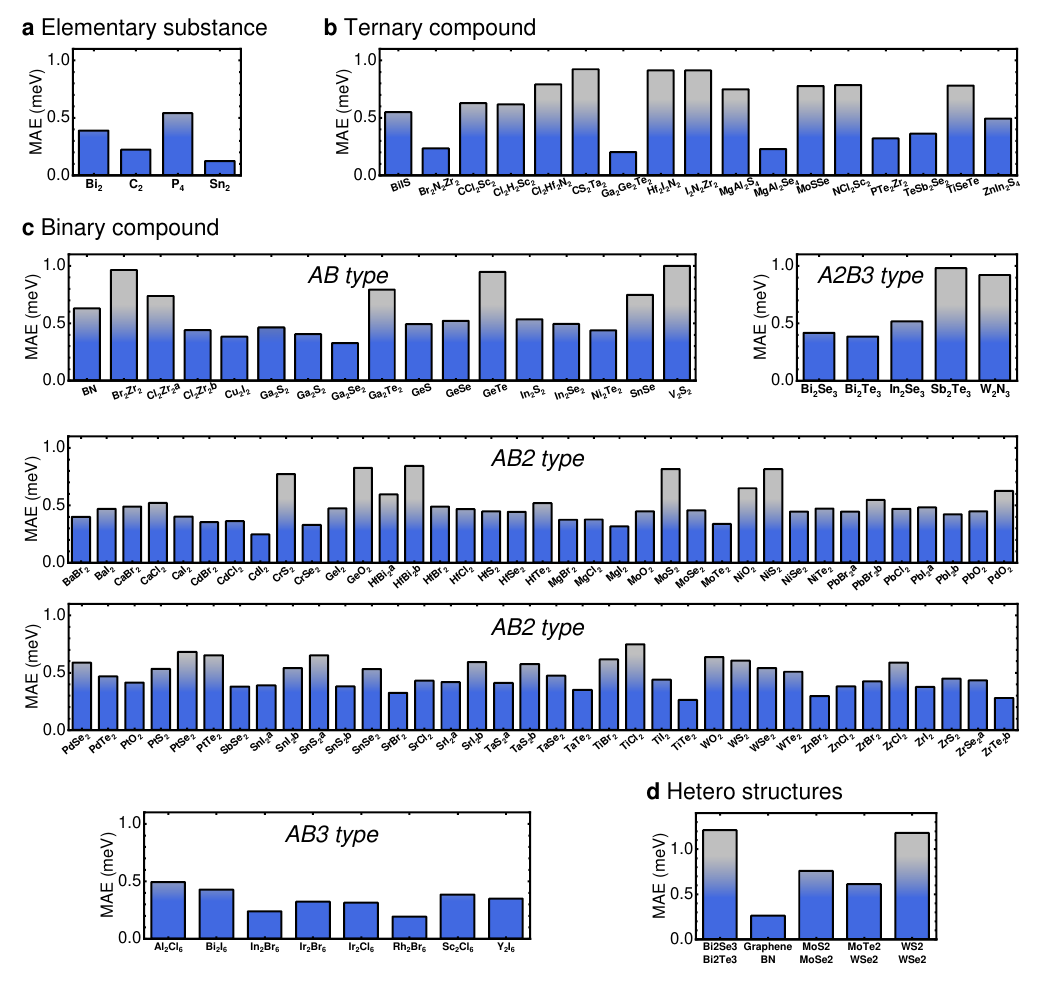}
    \caption{The statistics over MAEs between the DeepH-E3-predicted $H_\text{DeepH-E3}$ and the DFT-calculated $H_\text{DFT}$ of 124(5) types of twisted homobilayer(heterobilayer) materials covered in our database. (\textbf{a}-\textbf{d}) shows the MAEs of elementary substances, ternary compounds, binary compounds (divided into four prototypes \textit{AB}, \textit{$AB_2$}, \textit{$A_2B_3$} and \textit{$AB_3$}) and hetero structures, respectively, by bar plot. The height of each bar displays the MAE for one specific 2D material, which is obtained by averaging Hamiltonian errors over all atoms and orbitals for its twisted structures with the largest two commensurate twist angles $\theta =$ 21.79$^\circ$ (2-1) and 13.17$^\circ$ (3-2), except for $\theta =$ 10.87$^\circ$ (7-4) and 8.49$^\circ$ (9-5) in rectangular P4 (bP) and $\theta =$ 10.87$^\circ$ (7-4) and 8.49$^\circ$ (9-5) in square SnS$_2$. The predicted MAEs for most materials are in the order of sub meV, with only the twisted heterobilayer Bi$_2$Se$_3$/Bi$_2$Te$_3$ and WS$_2$/WSe$_2$ being exceptional (about 1.2 meV).}
    \label{fig:mae_heatmap}
\end{figure*}

\subsection{Accuracy of the Database}

To assess the accuracy and reliability of these DeepH-E3 models in our DDHT database in predicting Hamiltonians and electronic structures of twisted materials, we perform comprehensive statistics over mean absolute errors (MAEs) between the DeepH-E3-predicted and the DFT-calculated Hamiltonians for selected rigid-twisted structures of all 124 homo-bilayer materials. Considering the heavy computational demand of DFT Hamiltonians of twisted materials, we only cover the twisted supercells with either the largest two commensurate twist angles (e.g. 21.79$^\circ$ and 13.17$^\circ$ corresponding to twist index 2-1 and 3-2 for hexagonal lattice) or the number of atoms less than 120 in the MAE statistics. For each type of material, the absolute-error matrices are calculated by taking the absolute difference between respective hoppings of the predicted Hamiltonians of its twisted bilayer structures and that of the DFT calculated ones. These absolute error values are further averaged with respect to hopping pairs and containing chemical elements, resulting in the material-wise MAEs as shown in Fig. \ref{fig:mae_heatmap}. The overwhelming majority of hopping MAEs are in the sub-meV order of magnitude, which are comparable to or even less than commonly recognized errors in general DFT calculations. It's safe to say 1) the DeepH-E3 method can be efficiently generalized to much more materials using the workflow and parameter strategy here in the DDHT 2) Hamiltonian predictions by models in our DDHT database are indeed at DFT accuracy. By expanding the MAE heatmaps with respect to orbitals of different element pairs (\textcolor{blue}{see Note 5 in Supporting Information}), we find no significant large MAEs between these explicitly presented orbital-wise hoppings which may qualitatively weaken the above conclusion. 

It's also important to note that the models in our database work well for twisted materials both with and without spin-orbit coupling (SOC) in our demonstrations as long as the datasets (thus the DeepH-E3 models) incorporate SOC for respective (SOC-embedded) materials. 

\begin{figure*}
    \centering
    \includegraphics[width=\textwidth]{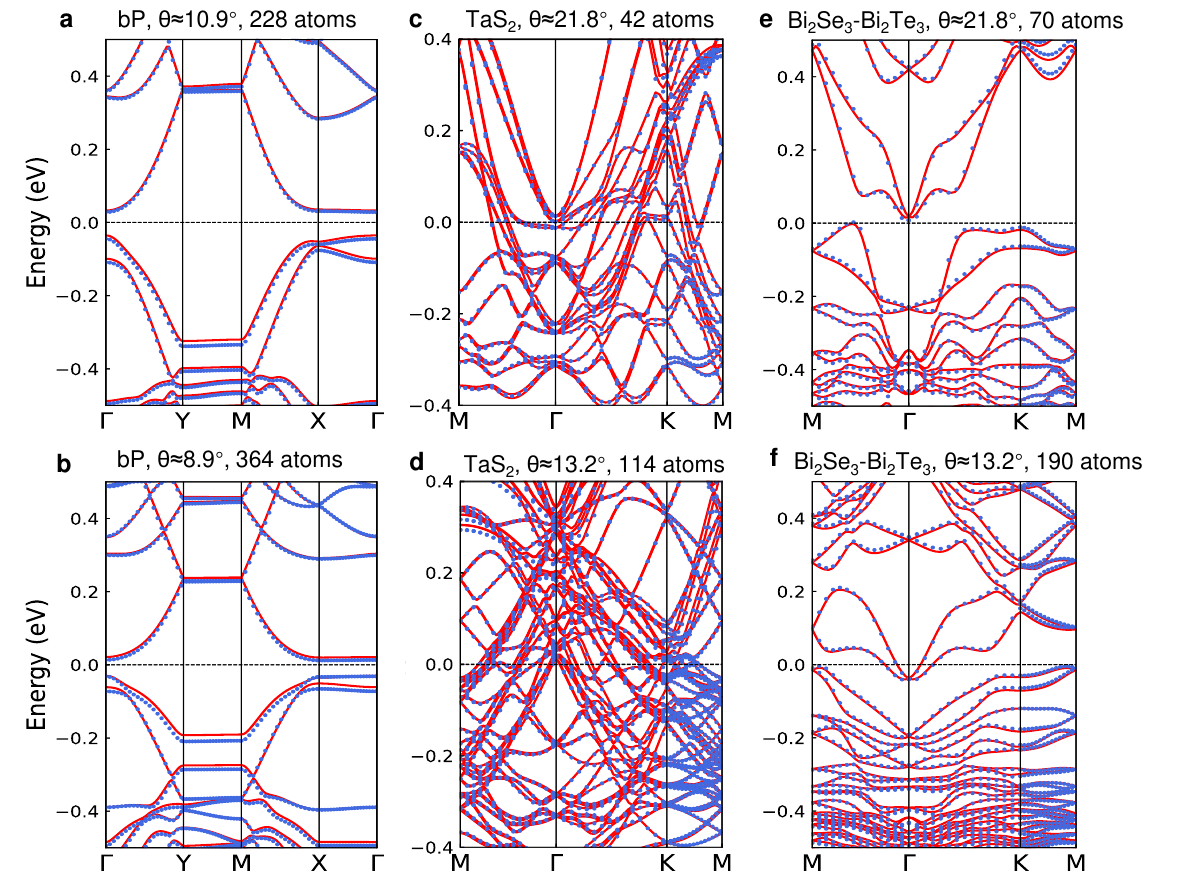}
    \caption{The DFT-calculated band structures (red curves) and predicted band structures (blue dots) for twisted (\textbf{a}, \textbf{b}) bP with $\theta=$ 10.87$^\circ$ and 8.49$^\circ$, and (\textbf{c}, \textbf{d}) 2H-phase TaS$_2$, and (\textbf{e}, \textbf{f}) heterobilayer Bi$_2$Se$_3$/Bi$_2$Te$_3$ with $\theta=$ 21.79$^\circ$ and 13.17$^\circ$. The twist angles of bP are different due to the consideration of lattice compression in constructing twisted supercells from rectangular lattices. The accuracy of both the DDHT models and predicted electronic structures is demonstrated by almost identical bands from predictions and DFT calculations.}
    \label{fig:band_dos}
\end{figure*}

We further verify the accuracy and effectiveness of the models in our database by calculating the electronic band structures from the above predicted Hamiltonian matrices and presenting the close fit between the predicted energy bands (energy dispersion relations) and the DFT-calculated ones. Here, two representative materials, bP and 2H-TaS$_2$, are chosen as demonstration based on the general research interest in their unique features, namely, the potential flat bands in twisted rectangular buckled lattice \cite{Kennes2020} of semiconducting bP and the interplay among twisting, SOC, charge density wave, and superconductivity in metallic TaS$_2$ \cite{Manzeli2017}. At large twist angles, the energy bands obtained from the DeepH-E3 predictions (blue dots) and the DFT calculations (red curves) are almost identical (left panels in Fig. \ref{fig:band_dos}a, b, d, e) in the energy range near the Fermi level. Similar results are observed in all other twisted homobilayer materials covered in our database, as shown in the left panels of Fig. SX and also reported in previous studies \cite{gong2023general,li_deep-learning_2022}. 

The twisted heterobilayers induce extra complexity in crystal structures and interatomic hopping relations, which is more challenging for DeepH-E3 models to give accurate predictions. We test the validity regarding the heterogeneity in their two constituent layers in several selected twisted heterobilayer materials, such as Bi$_2$Se$_3$/Bi$_2$Te$_3$, graphene/BN, MoS$_2$/MoSe$_2$, WS$_2$/WSe$_2$, and MoTe$_2$/WSe$_2$, and include the DeepH-E3 models of these materials in our database. Our database outputs accurate predictions for all these twisted heterobilayers considered (Fig. \ref{fig:band_dos}c, f and Fig. SXXX-SXXX) and therefore holds the accuracy and effectiveness toward heterogeneity in constituent layers. Take twisted strong-SOC Bi$_2$Se$_3$/Bi$_2$Te$_3$ with $\theta$ being 21.79$^\circ$ and 13.17$^\circ$ as an example, the predicted and DFT-calculated energy bands have negligible difference (with MAE in the order of XXX meV), and more importantly, the non-trivial Z2 topology of both twisted structures is also well reproduced by prediction. Therefore, as a step forward, our database firstly generalized the DeepH-E3 method and models to twisted heterobilayers with comparable accuracy to homobilayer cases. 


In addition to the DFT band structures, the predicted Hamiltonians by our database can be used to calculate the electron density of states (DOS) through identifying the basis projections of their eigenvectors. The right panels in Fig. \ref{fig:band_dos}a-f display the total (grey curves) and the projected DOS (colored curves) of twisted bP, 2H-TaS$_2$, and Bi$_2$Se$_3$/Bi$_2$Te$_3$, with DOS of all other large-angle twisted structures presented in the right panels of Fig. SX2. The predicted total and projected DOS provides more detailed information on the distribution of electronic states with respect to different energy levels, chemical elements, atoms, and orbitals, supplementing the picture of electronic structures other than dispersion relations demonstrated above. 

At the current stage, the above demonstrations of our database in accurately predicting DFT Hamiltonians and electronic structures cover all $124+5$ types of twisted bilayer materials considered, which indicates its high generalizability and reliability as a DFT-level deep-learning model database that offers a wide set of DeepH-E3 models targeting the prediction Hamiltonians of twisted materials, effectively bypassing the expensive first-principles calculations. The research into twisted materials and related areas can be greatly assisted and accelerated by our database, as well as the predicted results, through offering guidance and reference prior to costly investigations and screening for more exotic properties.

\subsection{Prediction of Flat Bands}

\begin{figure*}
     \centering
     \includegraphics[width=\textwidth]{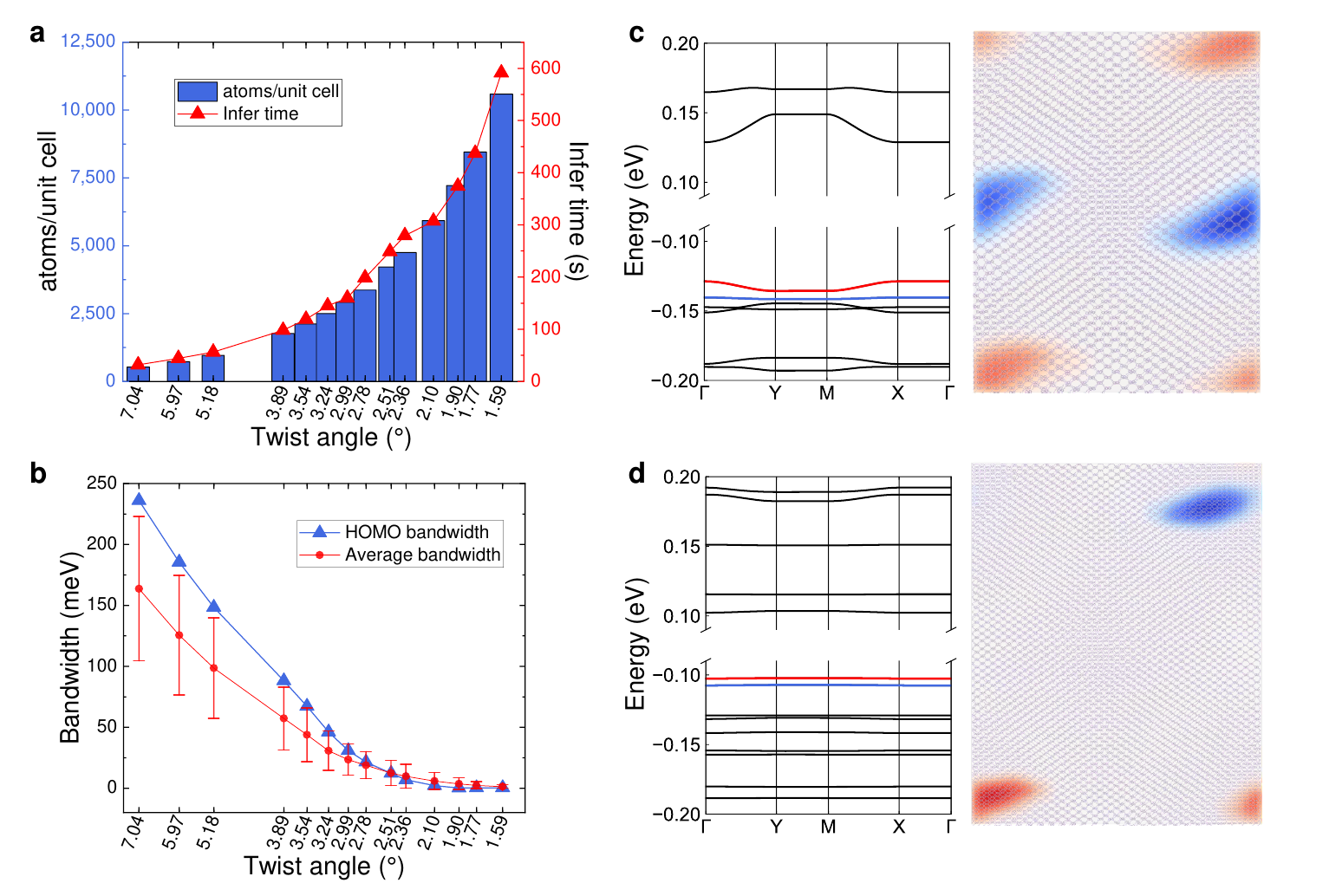}
     \caption{Prediction of flat bands in electronic structures of twisted bP systems. (\textbf{a}) shows the number of atoms per unit cell of twisted structures (blue bars) and the associated time of Hamiltonian inference using the trained model in DDHT for different twist angles. The inference time increases almost linearly with the number of atoms. (\textbf{b}) plots the bandwidth of highest occupied band, average bandwidth and standard error of 10 bands, which contains 5 highest valance bands and 5 lowest conduction bands, against the twist angle. The angle values on x-axis of (\textbf{a}, \textbf{b}) use logarithmic scale for clear demonstration. Predicted band structures and real-space flat-band projections of twisted bP structures with (\textbf{c}) $\theta=$ 2.36$^\circ$ and (\textbf{d}) 1.59$^\circ$ as examples. The top two flat valence bands are outlined by red and blue and their projections in real space are presented by transparent regions in the same color. The twisted atomic structures are schematically displayed in the background.}
     \label{fig:band_bp}
\end{figure*}

While our current demonstrations of predicting DFT Hamiltonians and electronic structures for twisted structures are limited to $\theta$ values as low as 13.17$^\circ$ for the majority of the considered materials, the remarkable generalizability of our database extends the capability to explore twisted structures with arbitrarily small $\theta$ angles—a domain beyond the computational reach of most DFT codes while maintaining acceptable computational costs. To illustrate this capability, we have chosen twisted bP bilayer structures at various twist angles as exemplars. As shown in the Fig. \ref{fig:band_dos}a, as the twist angle of bP approaches smaller values, the time required for the NN model to perform Hamiltonian inference, which is directly proportional to the number of atoms in the unit cell, increases rapidly but remains within the order of minutes. It is noteworthy that the widths of energy band near the Fermi energy diminish as the twist angle decreases, as shown in Fig. \ref{fig:band_dos}b, revealing the band width of the highest valence band and the average band width of the top 10 valence bands with respect to the twist angle of bP. The top two valence bands in Fig. \ref{fig:band_dos}a, d narrows their band width along the X-$\Gamma$ high-symmetry path with $\theta$ decreased by about 1.93$^\circ$, and they will be potentially flattened throughout the whole momentum space in twisted supercells at smaller $\theta$s. In this study, we examine two bP twisted supercells with $\theta=$ 2.36$^\circ$ (4756 atoms in unit cell) and 1.59$^\circ$ (10588 atoms in unit cell), which are relaxed by a machine-learning force-field model (GAP+R6) for phosphorene \cite{deringer2020general} to eliminate the unreasonable band crossings found in ion-clamped structures at small $\theta$s. Obtaining the band structures of the two bP twisted supercells with thousands of atoms using \textit{ab initio} DFT methods consumes unacceptable computational resources, however, they can be obtained using the DDHT model within days or even hours (mostly consumed by matrix diagonalization). As expected, the predicted band structures of the two small-angle bP twisted supercells, as shown in Fig. \ref{fig:band_bp}b, c, clearly display the correlated flat-band feature in their top valence states, offering a new opportunity to study flat-band physics and properties in an \textit{ab initio} manner. Obtaining the band structures of the two bP twisted structures with thousands of atoms using \textit{ab initio} DFT methods consumes unacceptable computational resources, however, they can be obtained using the DDHT model within days or even hours (mostly spent on matrix diagonalization).

For the twisted bilayer bP structure with twist angle $\theta=$ 2.36$^\circ$, our DDHT model for bP predicts the presence of two stand-alone flat bands (Fig. \ref{fig:band_bp}b) separated by 8.4 meV at the valence band top. The band width of the first (marked by red) and the second valence flat band (marked by blue) are 7.7 and 1.0 meV. Their small band width enables exotic correlated states when they are partially occupied upon hole doping. Moreover, the closely located flat bands with respect to energy make the correlated states from different flat bands accessible through modulation of doping density in a small range. The real-space projections of these two flat bands (\ref{fig:band_bp}e) are quite localized and well separated by half the moiré periodicity, which also support the strong correlated states. The flat bands and their real-space projections of twisted bP of smaller $\theta=$ 1.59$^\circ$ (Fig. \ref{fig:band_bp}c, f) are also predicted by our DDHT model. In Fig. \ref{fig:band_bp}c, multiple stand-alone flat valence bands are identified in energy range between $\pm$ 0.2 eV and have band width around 1.0 meV, potentially giving rise to more pronounced correlated phases than in larger-angle cases.

The accurate and efficient predictions of flat bands and their real-space projections pave the way for discovering new flat-band materials and future experimental and theoretical investigations into the correlated physics in twisted bP and other twisted materials. 


\section{Discussion}

While all the previous discussions have mainly focused on ion-clamped atomic structures of twisted materials, it should be noted that the DeepH-E3 models in the DDHT are not limited to predicting only such structures. In fact, our previous examples on the band-structure prediction of relaxed bP twisted bilayers demonstrated that the prediction accuracy is not significantly affected by structural relaxation. This is mainly because the atomic displacements induced by relaxation are generally below or around 0.2 \AA, the resulting interatomic distances between neighboring atoms after relaxation are still within the coverage of our training datasets (eg, Fig. S2). By using the same model that predicts for ion-clamped TBG, the DeepH-E3-predicted energy bands of relaxed magic-angle TBG fit closely to its DFT-calculated ones \cite{gong2023general}, which further support this argument. The obtained MAEs in both cases are in the sub-meV range, indicating that the models in the DDHT can capture the structural relaxation accurately. Therefore, we can conclude that the DDHT models are applicable to a wide range of twisted materials structures, whether they are ion-clamped or relaxed.

Straightforwardly, the models in the DDHT are highly versatile in predicting electronic structures of twisted multilayer structures, which can be treated as combinations of twisted bilayers consist of the same or different materials. For instance, in a twisted trilayer structure, where layers 1, 2, and 3 are arranged in a top-to-bottom order, layer 1-2 and layer 2-3 constitute two twisted bilayers. As a result, the Hamiltonian and electronic structure of this twisted trilayer can be accurately predicted using the same models as in the bilayer cases. This argument can be extended to twisted hetero-multilayer structures where the constituent layers are all different materials. In this case, multiple models trained by heterobilayer training datasets are required for accurate predictions. The accuracy of these models has been previously demonstrated by the predictions of twisted heterobilayer structures including graphene/BN, MoS$_2$/MoSe$_2$, WS$_2$/WSe$_2$, Bi$_2$Se$_3$/Bi$_2$Te$_3$, and MoTe$_2$/WSe$_2$, and can potentially sustain in predicting twisted multi-heterolayers.

The DDHT currently covers only a fraction of stable 2D materials (124 out of about 2132 non-magnetic materials in the C2DB accessed at June 19, 2022), it will cover more non-magnetic materials in the near-future development, by using the optimal scheme of dataset preparation and DeepH-E3 model training discussed in this article. Especially, we will take priority at offering models involving materials which users of our DDHT concern most. Although the models in the DDHT can predict arbitrary twist angles, the diagonalization to obtain band structure, DOS or other properties will consume certain amount of computational resource. Thus we allow users to upload their results help to enrich the database. In conclusion, the DDHT will be updated with predictions for more twist angles, especially for twisted materials and twist angles that are currently of general research interest. Additionally, the magnetic materials (about 464 stable in the C2DB) present a unique challenge due to their spin degree of freedom, which requires the incorporation of both spin moments and crystal structures as input in learning the neural network representations of their Hamiltonians. This can be achieved through the utilization of the recently developed xDeepH method \cite{xdeeph}, which can predict DFT Hamiltonians of magnetic twisted materials. Therefore, in the future development of the DDHT, xDeepH models and corresponding predictions of magnetic materials can be included. In our vision, the DDHT has the potential to cover all stable 2D materials in the future and serve as a handy tool for researchers in the emerging field of twistronics.

The predicted results in the DDHT are not limited to Hamiltonians and electronic structures, the identifications of topology in twisted materials will also be included in future development. This will be achieved by determining the irreducible representations of valence eigenstates at high-symmetry points in momentum space under various symmetry operations and comparing them with the elementary band representations (EBRs) in the topological quantum chemistry \cite{Bradlyn2017,Vergniory2019,Vergniory2022}. If these irreducible representations of a material are not linear combinations of EBRs, it would be identified as having non-trivial topology. Based on the predicted eigenstates by the DDHT models, theoretically, the band topology of twisted materials of any chemical constitutions and twist angles can be accurately and efficiently determined; meanwhile, the berry curvature may also be calculated to explore exotic properties regarding topology. Note that the classification of topology by topological quantum chemistry holds effective in the case of the generally-existed flat bands in twisted materials \cite{Clugru2021}. In this way, our DDHT may promisingly find all possible topological twisted materials and provide opportunities in exploring the currently scarce twist angle-dependent topological transitions.

\section{Summary}

In summary, by using the deep-learning method targeting at DFT Hamiltonians implemented in the DeepH-E3 code, we train the deep neural network models for 124(5) types of homo(hetero)bilayer 2D materials and use them to predict DFT Hamiltonians, as well as electronic structures, of their twisted structures at DFT accuracy. Based on these well-trained models capable of doing accurate predictions, we construct a deep-learning model database for Hamiltonian prediction of twisted materials named DDHT, which consists of DeepH-E3 NN models and related files like predicted DFT Hamiltonian matrices, eigenvalues and eigenvectors, band structures, density of states, and identified flat bands. Further, by screening the database, we list ultra flat bands in small-angle twisted bP and other 19 types of large-angle twisted materials. The accuracy of our DDHT models and predictions is demonstrated with comparable MAEs and energy bands to respective DFT results. Our DDHT covers twisted structures of a significant number of 2D materials and holds solid universality over a wide range of twist angles down to about 1$^\circ$, showing the reliability and universality of DeepH method and corresponding high-throughput NN parameter strategy, offering DFT-level knowledge prior to deep investigations and serving as a handy tool for researchers in the field of twistronics. Another crucial contribution of this work if to perform the new paradigm of building a material database, which is a collection of useful models applied to series of materials instead of single material and can be seen as a extensible MoE model. In the future development, our DDHT can be expanded to cover all stable 2D materials and incorporate magnetism and topology, satisfying the growing general interest toward exotic properties in twisted materials.

\section{Methods}
\subsection{Preparation of Training Datasets}
The bilayer unit cell for each type of 2D material in our database is fully relaxed using Viena \textit{ab initio} Simulation Package (VASP) \cite{kresse1996_1,PhysRevB.54.11169} to give the most optimal lattice constants, stacking orders, and interlayer spacings. The interactions between ions and electrons and the exchange correlations are treated by the projected-augmented wave method \cite{PhysRevB.50.17953,PhysRevB.59.1758} and the Perdew-Burke-Ernzerhof functional \cite{PhysRevLett.77.3865}. The interaction is corrected by DFT-D3 method \cite{RN1,RN2}. The convergence criteria for energy and force are $1\times10^{-7}$ eV and $1\times10^{-3}$ eV/\AA. The vacuum space between two periodic images along the direction normal to the 2D plane is set to be more than 12 \AA. The k-point mesh is 11$\times$11$\times$1 using the Monkhorst-Pack scheme \cite{PhysRevB.13.5188}.

The atomic structures in the training dataset for each 2D material in the DDHT are prepared as a series of $3\times3$ non-twisted bilayer supercells (made from the optimized primitive bilayer unit cells) with typically 24$\times$24 uniformly shifted interlayer stackings along two in-plane lattice vectors followed by random perturbations to atomic positions (up to 0.1 \AA) along three respective Cartesian axes. In total, the training dataset for each material has typically 576 perturbed and shifted $3\times3$ supercell structures. 

The Hamiltonian for each shifted and perturbed atomic structure in the training datasets is calculated by first-principles density functional theory (DFT) simulation code OpenMX \cite{Ozaki2003,Ozaki2004} using pseudo atomic orbital basis, norm-conserving pseudo potential \cite{Morrison1993}, and PBE exchange functional \cite{PhysRevLett.77.3865}. The energy convergence criteria, cut-off energy, and $\Gamma$-centered k-point mesh are set as $4\times10^{-8}$ Hartree, 300 Ry, and no less than $3\times3\times1$ in the self-consistent calculations, respectively. The DFT Hamiltonians and band structures used for evaluating the DDHT accuracy are calculated by OpenMX code using similar settings. The training datasets, including the DFT Hamiltonians and the corresponding structures, are processed prior to the training. The DFT Hamiltonians are decomposed into small matrix blocks $\mathcal{H}_{ij}$ that corresponds to local atomic structures $\mathcal{R}_{ij}$ within the cut-off radius $R_c$, and these $\mathcal{R}_{ij}$ are represented by crystal graphs where the on-site atoms and the distances to their neighbors within $R_c$ are treated by vertex and edge features. 

\subsection{Training and Prediction}

The deep neural network models in the DDHT are trained by the DeepH-E3 code \cite{li_deep-learning_2022,gong2023general} with Euclidean symmetry of local coordinate transformation satisfied by an E(3)-equivariant framework \cite{e3nn_paper}. Details of the DeepH-E3 code used in the DDHT are summarized as follows. The representation for the initial and the intermediate vertex (edge) features is 64$\times$0e and 64$\times$0e + 32$\times$1o + 16$\times$2e + 8$\times$3o + 8$\times$4e, respectively, where the first and the second number (separated by "$\times$") in each term (connected by "$+$") stands respectively for the number of vectors and the carried representation of spherical harmonics $l$, while the last letter e (o) denotes even (odd) parity. Spherical harmonics with $l=0$ to 4 and $l=0$ to 5 are used for materials without and with spin-orbit coupling respectively. During the model training, we employ an optimized learning scheme where the training, validation, and testing set take respectively 80\%, 15\%, and 5\% of the total dataset, the batch size takes the value of 2, and the learning rate starts with 0.002 and reduces by half when validation loss plateau (in this case, decreasing less than 5\% after every 120 epochs). The training takes at least 1,000 epochs and is terminated upon the validation loss drops below $1\times10^{-7}$ for its averaged value or $1\times10^{-5}$ for its maximum value, ensuring the high accuracy and generalizability of DeepH-E3 models. Similar loss values are given in evaluation of these models in testing sets upon the stopping of model training. 

The Hamiltonian matrices are predicted using only atomic structures of twisted bilayer materials, corresponding overlap matrices, and trained DeepH-E3 models as input. Based on the atomic structures and local-atomic basis, the overlap matrices are directly obtained by a modified OpenMX code without conducting DFT calculations. The $H_{\text{DeepH-E3}}$ matrices are sparse due to the locality of local-atomic basis and the eigenvalues and eigenvectors are calculated by efficient matrix diagonalization using the Pardiso package \cite{pardiso-8.0a, pardiso-8.0c} of Julia code. 

\section{Data Availability}
The DeepH-E3 models of all 124 homobilayer and 5 heterobilayer materials and their twisted structures of large $\theta$s will be available at the DDHT website.

\section{Acknowledgments}
This work was supported by the Basic Science Center Project of NSFC (grant no. 2388201), the National Natural Science Foundation of China (grant no. 12334003), the National Science Fund for Distinguished Young Scholars (grant no. 12025405) and the Ministry of Science and Technology of China (grant no. 2023YFA1406400). R.X. acknowledges the founding support from the Fellowship of China Postdoctoral Science Foundation (Grant No. 2021TQ0187).
to be updated

\section{Author Contributions}
T.B. and R.X. contributed equally to this project. 

\section{Competing Interests}
The authors declare no competing of interests.

\bibliography{ref}

\begin{thebibliography}{50}%
\makeatletter
\providecommand \@ifxundefined [1]{%
 \@ifx{#1\undefined}
}%
\providecommand \@ifnum [1]{%
 \ifnum #1\expandafter \@firstoftwo
 \else \expandafter \@secondoftwo
 \fi
}%
\providecommand \@ifx [1]{%
 \ifx #1\expandafter \@firstoftwo
 \else \expandafter \@secondoftwo
 \fi
}%
\providecommand \natexlab [1]{#1}%
\providecommand \enquote  [1]{``#1''}%
\providecommand \bibnamefont  [1]{#1}%
\providecommand \bibfnamefont [1]{#1}%
\providecommand \citenamefont [1]{#1}%
\providecommand \href@noop [0]{\@secondoftwo}%
\providecommand \href [0]{\begingroup \@sanitize@url \@href}%
\providecommand \@href[1]{\@@startlink{#1}\@@href}%
\providecommand \@@href[1]{\endgroup#1\@@endlink}%
\providecommand \@sanitize@url [0]{\catcode `\\12\catcode `\$12\catcode
  `\&12\catcode `\#12\catcode `\^12\catcode `\_12\catcode `\%12\relax}%
\providecommand \@@startlink[1]{}%
\providecommand \@@endlink[0]{}%
\providecommand \url  [0]{\begingroup\@sanitize@url \@url }%
\providecommand \@url [1]{\endgroup\@href {#1}{\urlprefix }}%
\providecommand \urlprefix  [0]{URL }%
\providecommand \Eprint [0]{\href }%
\providecommand \doibase [0]{https://doi.org/}%
\providecommand \selectlanguage [0]{\@gobble}%
\providecommand \bibinfo  [0]{\@secondoftwo}%
\providecommand \bibfield  [0]{\@secondoftwo}%
\providecommand \translation [1]{[#1]}%
\providecommand \BibitemOpen [0]{}%
\providecommand \bibitemStop [0]{}%
\providecommand \bibitemNoStop [0]{.\EOS\space}%
\providecommand \EOS [0]{\spacefactor3000\relax}%
\providecommand \BibitemShut  [1]{\csname bibitem#1\endcsname}%
\let\auto@bib@innerbib\@empty
\bibitem [{\citenamefont {Cao}\ \emph {et~al.}(2018{\natexlab{a}})\citenamefont
  {Cao}, \citenamefont {Fatemi}, \citenamefont {Demir}, \citenamefont {Fang},
  \citenamefont {Tomarken}, \citenamefont {Luo}, \citenamefont
  {Sanchez-Yamagishi}, \citenamefont {Watanabe}, \citenamefont {Taniguchi},
  \citenamefont {Kaxiras}, \citenamefont {Ashoori},\ and\ \citenamefont
  {Jarillo-Herrero}}]{Cao2018_1}%
  \BibitemOpen
  \bibfield  {author} {\bibinfo {author} {\bibfnamefont {Y.}~\bibnamefont
  {Cao}}, \bibinfo {author} {\bibfnamefont {V.}~\bibnamefont {Fatemi}},
  \bibinfo {author} {\bibfnamefont {A.}~\bibnamefont {Demir}}, \bibinfo
  {author} {\bibfnamefont {S.}~\bibnamefont {Fang}}, \bibinfo {author}
  {\bibfnamefont {S.~L.}\ \bibnamefont {Tomarken}}, \bibinfo {author}
  {\bibfnamefont {J.~Y.}\ \bibnamefont {Luo}}, \bibinfo {author} {\bibfnamefont
  {J.~D.}\ \bibnamefont {Sanchez-Yamagishi}}, \bibinfo {author} {\bibfnamefont
  {K.}~\bibnamefont {Watanabe}}, \bibinfo {author} {\bibfnamefont
  {T.}~\bibnamefont {Taniguchi}}, \bibinfo {author} {\bibfnamefont
  {E.}~\bibnamefont {Kaxiras}}, \bibinfo {author} {\bibfnamefont {R.~C.}\
  \bibnamefont {Ashoori}},\ and\ \bibinfo {author} {\bibfnamefont
  {P.}~\bibnamefont {Jarillo-Herrero}},\ }\bibfield  {title} {\bibinfo {title}
  {Correlated insulator behaviour at half-filling in magic-angle graphene
  superlattices},\ }\href {https://doi.org/10.1038/nature26154} {\bibfield
  {journal} {\bibinfo  {journal} {Nature}\ }\textbf {\bibinfo {volume} {556}},\
  \bibinfo {pages} {80} (\bibinfo {year} {2018}{\natexlab{a}})}\BibitemShut
  {NoStop}%
\bibitem [{\citenamefont {Cao}\ \emph {et~al.}(2018{\natexlab{b}})\citenamefont
  {Cao}, \citenamefont {Fatemi}, \citenamefont {Fang}, \citenamefont
  {Watanabe}, \citenamefont {Taniguchi}, \citenamefont {Kaxiras},\ and\
  \citenamefont {Jarillo-Herrero}}]{Cao2018_2}%
  \BibitemOpen
  \bibfield  {author} {\bibinfo {author} {\bibfnamefont {Y.}~\bibnamefont
  {Cao}}, \bibinfo {author} {\bibfnamefont {V.}~\bibnamefont {Fatemi}},
  \bibinfo {author} {\bibfnamefont {S.}~\bibnamefont {Fang}}, \bibinfo {author}
  {\bibfnamefont {K.}~\bibnamefont {Watanabe}}, \bibinfo {author}
  {\bibfnamefont {T.}~\bibnamefont {Taniguchi}}, \bibinfo {author}
  {\bibfnamefont {E.}~\bibnamefont {Kaxiras}},\ and\ \bibinfo {author}
  {\bibfnamefont {P.}~\bibnamefont {Jarillo-Herrero}},\ }\bibfield  {title}
  {\bibinfo {title} {Unconventional superconductivity in magic-angle graphene
  superlattices},\ }\href {https://doi.org/10.1038/nature26160} {\bibfield
  {journal} {\bibinfo  {journal} {Nature}\ }\textbf {\bibinfo {volume} {556}},\
  \bibinfo {pages} {43} (\bibinfo {year} {2018}{\natexlab{b}})}\BibitemShut
  {NoStop}%
\bibitem [{\citenamefont {Sun}\ \emph {et~al.}(2024)\citenamefont {Sun},
  \citenamefont {Suriyage}, \citenamefont {Khan}, \citenamefont {Gao},
  \citenamefont {Zhao}, \citenamefont {Liu}, \citenamefont {Hasan},
  \citenamefont {Rahman}, \citenamefont {Chen}, \citenamefont {Lam} \emph
  {et~al.}}]{sun2024twisted}%
  \BibitemOpen
  \bibfield  {author} {\bibinfo {author} {\bibfnamefont {X.}~\bibnamefont
  {Sun}}, \bibinfo {author} {\bibfnamefont {M.}~\bibnamefont {Suriyage}},
  \bibinfo {author} {\bibfnamefont {A.~R.}\ \bibnamefont {Khan}}, \bibinfo
  {author} {\bibfnamefont {M.}~\bibnamefont {Gao}}, \bibinfo {author}
  {\bibfnamefont {J.}~\bibnamefont {Zhao}}, \bibinfo {author} {\bibfnamefont
  {B.}~\bibnamefont {Liu}}, \bibinfo {author} {\bibfnamefont {M.~M.}\
  \bibnamefont {Hasan}}, \bibinfo {author} {\bibfnamefont {S.}~\bibnamefont
  {Rahman}}, \bibinfo {author} {\bibfnamefont {R.-s.}\ \bibnamefont {Chen}},
  \bibinfo {author} {\bibfnamefont {P.~K.}\ \bibnamefont {Lam}}, \emph
  {et~al.},\ }\bibfield  {title} {\bibinfo {title} {Twisted van der waals
  quantum materials: Fundamentals, tunability, and applications},\ }\href@noop
  {} {\bibfield  {journal} {\bibinfo  {journal} {Chemical Reviews}\ } (\bibinfo
  {year} {2024})}\BibitemShut {NoStop}%
\bibitem [{\citenamefont {Song}\ \emph {et~al.}(2021)\citenamefont {Song},
  \citenamefont {Sun}, \citenamefont {Anderson}, \citenamefont {Wang},
  \citenamefont {Qian}, \citenamefont {Taniguchi}, \citenamefont {Watanabe},
  \citenamefont {McGuire}, \citenamefont {St\"{o}hr}, \citenamefont {Xiao},
  \citenamefont {Cao}, \citenamefont {Wrachtrup},\ and\ \citenamefont
  {Xu}}]{Song2021}%
  \BibitemOpen
  \bibfield  {author} {\bibinfo {author} {\bibfnamefont {T.}~\bibnamefont
  {Song}}, \bibinfo {author} {\bibfnamefont {Q.-C.}\ \bibnamefont {Sun}},
  \bibinfo {author} {\bibfnamefont {E.}~\bibnamefont {Anderson}}, \bibinfo
  {author} {\bibfnamefont {C.}~\bibnamefont {Wang}}, \bibinfo {author}
  {\bibfnamefont {J.}~\bibnamefont {Qian}}, \bibinfo {author} {\bibfnamefont
  {T.}~\bibnamefont {Taniguchi}}, \bibinfo {author} {\bibfnamefont
  {K.}~\bibnamefont {Watanabe}}, \bibinfo {author} {\bibfnamefont {M.~A.}\
  \bibnamefont {McGuire}}, \bibinfo {author} {\bibfnamefont {R.}~\bibnamefont
  {St\"{o}hr}}, \bibinfo {author} {\bibfnamefont {D.}~\bibnamefont {Xiao}},
  \bibinfo {author} {\bibfnamefont {T.}~\bibnamefont {Cao}}, \bibinfo {author}
  {\bibfnamefont {J.}~\bibnamefont {Wrachtrup}},\ and\ \bibinfo {author}
  {\bibfnamefont {X.}~\bibnamefont {Xu}},\ }\bibfield  {title} {\bibinfo
  {title} {Direct visualization of magnetic domains and moir{\'{e}} magnetism
  in twisted 2d magnets},\ }\href {https://doi.org/10.1126/science.abj7478}
  {\bibfield  {journal} {\bibinfo  {journal} {Science}\ }\textbf {\bibinfo
  {volume} {374}},\ \bibinfo {pages} {1140} (\bibinfo {year}
  {2021})}\BibitemShut {NoStop}%
\bibitem [{\citenamefont {Xu}\ \emph {et~al.}(2021)\citenamefont {Xu},
  \citenamefont {Ray}, \citenamefont {Shao}, \citenamefont {Jiang},
  \citenamefont {Lee}, \citenamefont {Weber}, \citenamefont {Goldberger},
  \citenamefont {Watanabe}, \citenamefont {Taniguchi}, \citenamefont {Muller},
  \citenamefont {Mak},\ and\ \citenamefont {Shan}}]{Xu2021}%
  \BibitemOpen
  \bibfield  {author} {\bibinfo {author} {\bibfnamefont {Y.}~\bibnamefont
  {Xu}}, \bibinfo {author} {\bibfnamefont {A.}~\bibnamefont {Ray}}, \bibinfo
  {author} {\bibfnamefont {Y.-T.}\ \bibnamefont {Shao}}, \bibinfo {author}
  {\bibfnamefont {S.}~\bibnamefont {Jiang}}, \bibinfo {author} {\bibfnamefont
  {K.}~\bibnamefont {Lee}}, \bibinfo {author} {\bibfnamefont {D.}~\bibnamefont
  {Weber}}, \bibinfo {author} {\bibfnamefont {J.~E.}\ \bibnamefont
  {Goldberger}}, \bibinfo {author} {\bibfnamefont {K.}~\bibnamefont
  {Watanabe}}, \bibinfo {author} {\bibfnamefont {T.}~\bibnamefont {Taniguchi}},
  \bibinfo {author} {\bibfnamefont {D.~A.}\ \bibnamefont {Muller}}, \bibinfo
  {author} {\bibfnamefont {K.~F.}\ \bibnamefont {Mak}},\ and\ \bibinfo {author}
  {\bibfnamefont {J.}~\bibnamefont {Shan}},\ }\bibfield  {title} {\bibinfo
  {title} {Coexisting ferromagnetic{\textendash}antiferromagnetic state in
  twisted bilayer {CrI}3},\ }\href {https://doi.org/10.1038/s41565-021-01014-y}
  {\bibfield  {journal} {\bibinfo  {journal} {Nature Nanotechnology}\ }\textbf
  {\bibinfo {volume} {17}},\ \bibinfo {pages} {143} (\bibinfo {year}
  {2021})}\BibitemShut {NoStop}%
\bibitem [{\citenamefont {Stern}\ \emph {et~al.}(2021)\citenamefont {Stern},
  \citenamefont {Waschitz}, \citenamefont {Cao}, \citenamefont {Nevo},
  \citenamefont {Watanabe}, \citenamefont {Taniguchi}, \citenamefont {Sela},
  \citenamefont {Urbakh}, \citenamefont {Hod},\ and\ \citenamefont
  {Shalom}}]{ViznerStern2021}%
  \BibitemOpen
  \bibfield  {author} {\bibinfo {author} {\bibfnamefont {M.~V.}\ \bibnamefont
  {Stern}}, \bibinfo {author} {\bibfnamefont {Y.}~\bibnamefont {Waschitz}},
  \bibinfo {author} {\bibfnamefont {W.}~\bibnamefont {Cao}}, \bibinfo {author}
  {\bibfnamefont {I.}~\bibnamefont {Nevo}}, \bibinfo {author} {\bibfnamefont
  {K.}~\bibnamefont {Watanabe}}, \bibinfo {author} {\bibfnamefont
  {T.}~\bibnamefont {Taniguchi}}, \bibinfo {author} {\bibfnamefont
  {E.}~\bibnamefont {Sela}}, \bibinfo {author} {\bibfnamefont {M.}~\bibnamefont
  {Urbakh}}, \bibinfo {author} {\bibfnamefont {O.}~\bibnamefont {Hod}},\ and\
  \bibinfo {author} {\bibfnamefont {M.~B.}\ \bibnamefont {Shalom}},\ }\bibfield
   {title} {\bibinfo {title} {Interfacial ferroelectricity by van der waals
  sliding},\ }\href {https://doi.org/10.1126/science.abe8177} {\bibfield
  {journal} {\bibinfo  {journal} {Science}\ }\textbf {\bibinfo {volume}
  {372}},\ \bibinfo {pages} {1462} (\bibinfo {year} {2021})}\BibitemShut
  {NoStop}%
\bibitem [{\citenamefont {Yasuda}\ \emph {et~al.}(2021)\citenamefont {Yasuda},
  \citenamefont {Wang}, \citenamefont {Watanabe}, \citenamefont {Taniguchi},\
  and\ \citenamefont {Jarillo-Herrero}}]{Yasuda2021}%
  \BibitemOpen
  \bibfield  {author} {\bibinfo {author} {\bibfnamefont {K.}~\bibnamefont
  {Yasuda}}, \bibinfo {author} {\bibfnamefont {X.}~\bibnamefont {Wang}},
  \bibinfo {author} {\bibfnamefont {K.}~\bibnamefont {Watanabe}}, \bibinfo
  {author} {\bibfnamefont {T.}~\bibnamefont {Taniguchi}},\ and\ \bibinfo
  {author} {\bibfnamefont {P.}~\bibnamefont {Jarillo-Herrero}},\ }\bibfield
  {title} {\bibinfo {title} {Stacking-engineered ferroelectricity in bilayer
  boron nitride},\ }\href {https://doi.org/10.1126/science.abd3230} {\bibfield
  {journal} {\bibinfo  {journal} {Science}\ }\textbf {\bibinfo {volume}
  {372}},\ \bibinfo {pages} {1458} (\bibinfo {year} {2021})}\BibitemShut
  {NoStop}%
\bibitem [{\citenamefont {Devakul}\ \emph {et~al.}(2021)\citenamefont
  {Devakul}, \citenamefont {Cr{\'e}pel}, \citenamefont {Zhang},\ and\
  \citenamefont {Fu}}]{devakul2021magic}%
  \BibitemOpen
  \bibfield  {author} {\bibinfo {author} {\bibfnamefont {T.}~\bibnamefont
  {Devakul}}, \bibinfo {author} {\bibfnamefont {V.}~\bibnamefont {Cr{\'e}pel}},
  \bibinfo {author} {\bibfnamefont {Y.}~\bibnamefont {Zhang}},\ and\ \bibinfo
  {author} {\bibfnamefont {L.}~\bibnamefont {Fu}},\ }\bibfield  {title}
  {\bibinfo {title} {Magic in twisted transition metal dichalcogenide
  bilayers},\ }\href@noop {} {\bibfield  {journal} {\bibinfo  {journal} {Nature
  communications}\ }\textbf {\bibinfo {volume} {12}},\ \bibinfo {pages} {6730}
  (\bibinfo {year} {2021})}\BibitemShut {NoStop}%
\bibitem [{\citenamefont {Jin}\ \emph {et~al.}(2019)\citenamefont {Jin},
  \citenamefont {Regan}, \citenamefont {Yan}, \citenamefont {Utama},
  \citenamefont {Wang}, \citenamefont {Zhao}, \citenamefont {Qin},
  \citenamefont {Yang}, \citenamefont {Zheng}, \citenamefont {Shi},
  \citenamefont {Watanabe}, \citenamefont {Taniguchi}, \citenamefont {Tongay},
  \citenamefont {Zettl},\ and\ \citenamefont {Wang}}]{Jin2019}%
  \BibitemOpen
  \bibfield  {author} {\bibinfo {author} {\bibfnamefont {C.}~\bibnamefont
  {Jin}}, \bibinfo {author} {\bibfnamefont {E.~C.}\ \bibnamefont {Regan}},
  \bibinfo {author} {\bibfnamefont {A.}~\bibnamefont {Yan}}, \bibinfo {author}
  {\bibfnamefont {M.~I.~B.}\ \bibnamefont {Utama}}, \bibinfo {author}
  {\bibfnamefont {D.}~\bibnamefont {Wang}}, \bibinfo {author} {\bibfnamefont
  {S.}~\bibnamefont {Zhao}}, \bibinfo {author} {\bibfnamefont {Y.}~\bibnamefont
  {Qin}}, \bibinfo {author} {\bibfnamefont {S.}~\bibnamefont {Yang}}, \bibinfo
  {author} {\bibfnamefont {Z.}~\bibnamefont {Zheng}}, \bibinfo {author}
  {\bibfnamefont {S.}~\bibnamefont {Shi}}, \bibinfo {author} {\bibfnamefont
  {K.}~\bibnamefont {Watanabe}}, \bibinfo {author} {\bibfnamefont
  {T.}~\bibnamefont {Taniguchi}}, \bibinfo {author} {\bibfnamefont
  {S.}~\bibnamefont {Tongay}}, \bibinfo {author} {\bibfnamefont
  {A.}~\bibnamefont {Zettl}},\ and\ \bibinfo {author} {\bibfnamefont
  {F.}~\bibnamefont {Wang}},\ }\bibfield  {title} {\bibinfo {title}
  {Observation of moir{\'{e}} excitons in {WSe}2/{WS}2 heterostructure
  superlattices},\ }\href {https://doi.org/10.1038/s41586-019-0976-y}
  {\bibfield  {journal} {\bibinfo  {journal} {Nature}\ }\textbf {\bibinfo
  {volume} {567}},\ \bibinfo {pages} {76} (\bibinfo {year} {2019})}\BibitemShut
  {NoStop}%
\bibitem [{\citenamefont {Tran}\ \emph {et~al.}(2019)\citenamefont {Tran},
  \citenamefont {Moody}, \citenamefont {Wu}, \citenamefont {Lu}, \citenamefont
  {Choi}, \citenamefont {Kim}, \citenamefont {Rai}, \citenamefont {Sanchez},
  \citenamefont {Quan}, \citenamefont {Singh}, \citenamefont {Embley},
  \citenamefont {Zepeda}, \citenamefont {Campbell}, \citenamefont {Autry},
  \citenamefont {Taniguchi}, \citenamefont {Watanabe}, \citenamefont {Lu},
  \citenamefont {Banerjee}, \citenamefont {Silverman}, \citenamefont {Kim},
  \citenamefont {Tutuc}, \citenamefont {Yang}, \citenamefont {MacDonald},\ and\
  \citenamefont {Li}}]{Tran2019}%
  \BibitemOpen
  \bibfield  {author} {\bibinfo {author} {\bibfnamefont {K.}~\bibnamefont
  {Tran}}, \bibinfo {author} {\bibfnamefont {G.}~\bibnamefont {Moody}},
  \bibinfo {author} {\bibfnamefont {F.}~\bibnamefont {Wu}}, \bibinfo {author}
  {\bibfnamefont {X.}~\bibnamefont {Lu}}, \bibinfo {author} {\bibfnamefont
  {J.}~\bibnamefont {Choi}}, \bibinfo {author} {\bibfnamefont {K.}~\bibnamefont
  {Kim}}, \bibinfo {author} {\bibfnamefont {A.}~\bibnamefont {Rai}}, \bibinfo
  {author} {\bibfnamefont {D.~A.}\ \bibnamefont {Sanchez}}, \bibinfo {author}
  {\bibfnamefont {J.}~\bibnamefont {Quan}}, \bibinfo {author} {\bibfnamefont
  {A.}~\bibnamefont {Singh}}, \bibinfo {author} {\bibfnamefont
  {J.}~\bibnamefont {Embley}}, \bibinfo {author} {\bibfnamefont
  {A.}~\bibnamefont {Zepeda}}, \bibinfo {author} {\bibfnamefont
  {M.}~\bibnamefont {Campbell}}, \bibinfo {author} {\bibfnamefont
  {T.}~\bibnamefont {Autry}}, \bibinfo {author} {\bibfnamefont
  {T.}~\bibnamefont {Taniguchi}}, \bibinfo {author} {\bibfnamefont
  {K.}~\bibnamefont {Watanabe}}, \bibinfo {author} {\bibfnamefont
  {N.}~\bibnamefont {Lu}}, \bibinfo {author} {\bibfnamefont {S.~K.}\
  \bibnamefont {Banerjee}}, \bibinfo {author} {\bibfnamefont {K.~L.}\
  \bibnamefont {Silverman}}, \bibinfo {author} {\bibfnamefont {S.}~\bibnamefont
  {Kim}}, \bibinfo {author} {\bibfnamefont {E.}~\bibnamefont {Tutuc}}, \bibinfo
  {author} {\bibfnamefont {L.}~\bibnamefont {Yang}}, \bibinfo {author}
  {\bibfnamefont {A.~H.}\ \bibnamefont {MacDonald}},\ and\ \bibinfo {author}
  {\bibfnamefont {X.}~\bibnamefont {Li}},\ }\bibfield  {title} {\bibinfo
  {title} {Evidence for moir{\'{e}} excitons in van der waals
  heterostructures},\ }\href {https://doi.org/10.1038/s41586-019-0975-z}
  {\bibfield  {journal} {\bibinfo  {journal} {Nature}\ }\textbf {\bibinfo
  {volume} {567}},\ \bibinfo {pages} {71} (\bibinfo {year} {2019})}\BibitemShut
  {NoStop}%
\bibitem [{\citenamefont {Alexeev}\ \emph {et~al.}(2019)\citenamefont
  {Alexeev}, \citenamefont {Ruiz-Tijerina}, \citenamefont {Danovich},
  \citenamefont {Hamer}, \citenamefont {Terry}, \citenamefont {Nayak},
  \citenamefont {Ahn}, \citenamefont {Pak}, \citenamefont {Lee}, \citenamefont
  {Sohn}, \citenamefont {Molas}, \citenamefont {Koperski}, \citenamefont
  {Watanabe}, \citenamefont {Taniguchi}, \citenamefont {Novoselov},
  \citenamefont {Gorbachev}, \citenamefont {Shin}, \citenamefont {Fal'ko},\
  and\ \citenamefont {Tartakovskii}}]{Alexeev2019}%
  \BibitemOpen
  \bibfield  {author} {\bibinfo {author} {\bibfnamefont {E.~M.}\ \bibnamefont
  {Alexeev}}, \bibinfo {author} {\bibfnamefont {D.~A.}\ \bibnamefont
  {Ruiz-Tijerina}}, \bibinfo {author} {\bibfnamefont {M.}~\bibnamefont
  {Danovich}}, \bibinfo {author} {\bibfnamefont {M.~J.}\ \bibnamefont {Hamer}},
  \bibinfo {author} {\bibfnamefont {D.~J.}\ \bibnamefont {Terry}}, \bibinfo
  {author} {\bibfnamefont {P.~K.}\ \bibnamefont {Nayak}}, \bibinfo {author}
  {\bibfnamefont {S.}~\bibnamefont {Ahn}}, \bibinfo {author} {\bibfnamefont
  {S.}~\bibnamefont {Pak}}, \bibinfo {author} {\bibfnamefont {J.}~\bibnamefont
  {Lee}}, \bibinfo {author} {\bibfnamefont {J.~I.}\ \bibnamefont {Sohn}},
  \bibinfo {author} {\bibfnamefont {M.~R.}\ \bibnamefont {Molas}}, \bibinfo
  {author} {\bibfnamefont {M.}~\bibnamefont {Koperski}}, \bibinfo {author}
  {\bibfnamefont {K.}~\bibnamefont {Watanabe}}, \bibinfo {author}
  {\bibfnamefont {T.}~\bibnamefont {Taniguchi}}, \bibinfo {author}
  {\bibfnamefont {K.~S.}\ \bibnamefont {Novoselov}}, \bibinfo {author}
  {\bibfnamefont {R.~V.}\ \bibnamefont {Gorbachev}}, \bibinfo {author}
  {\bibfnamefont {H.~S.}\ \bibnamefont {Shin}}, \bibinfo {author}
  {\bibfnamefont {V.~I.}\ \bibnamefont {Fal'ko}},\ and\ \bibinfo {author}
  {\bibfnamefont {A.~I.}\ \bibnamefont {Tartakovskii}},\ }\bibfield  {title}
  {\bibinfo {title} {Resonantly hybridized excitons in moir{\'{e}}
  superlattices in van der waals heterostructures},\ }\href
  {https://doi.org/10.1038/s41586-019-0986-9} {\bibfield  {journal} {\bibinfo
  {journal} {Nature}\ }\textbf {\bibinfo {volume} {567}},\ \bibinfo {pages}
  {81} (\bibinfo {year} {2019})}\BibitemShut {NoStop}%
\bibitem [{\citenamefont {Shimazaki}\ \emph {et~al.}(2020)\citenamefont
  {Shimazaki}, \citenamefont {Schwartz}, \citenamefont {Watanabe},
  \citenamefont {Taniguchi}, \citenamefont {Kroner},\ and\ \citenamefont
  {Imamo{\u{g}}lu}}]{Shimazaki2020}%
  \BibitemOpen
  \bibfield  {author} {\bibinfo {author} {\bibfnamefont {Y.}~\bibnamefont
  {Shimazaki}}, \bibinfo {author} {\bibfnamefont {I.}~\bibnamefont {Schwartz}},
  \bibinfo {author} {\bibfnamefont {K.}~\bibnamefont {Watanabe}}, \bibinfo
  {author} {\bibfnamefont {T.}~\bibnamefont {Taniguchi}}, \bibinfo {author}
  {\bibfnamefont {M.}~\bibnamefont {Kroner}},\ and\ \bibinfo {author}
  {\bibfnamefont {A.}~\bibnamefont {Imamo{\u{g}}lu}},\ }\bibfield  {title}
  {\bibinfo {title} {Strongly correlated electrons and hybrid excitons in a
  moir{\'{e}} heterostructure},\ }\href
  {https://doi.org/10.1038/s41586-020-2191-2} {\bibfield  {journal} {\bibinfo
  {journal} {Nature}\ }\textbf {\bibinfo {volume} {580}},\ \bibinfo {pages}
  {472} (\bibinfo {year} {2020})}\BibitemShut {NoStop}%
\bibitem [{\citenamefont {Regan}\ \emph {et~al.}(2020)\citenamefont {Regan},
  \citenamefont {Wang}, \citenamefont {Jin}, \citenamefont {Utama},
  \citenamefont {Gao}, \citenamefont {Wei}, \citenamefont {Zhao}, \citenamefont
  {Zhao}, \citenamefont {Zhang}, \citenamefont {Yumigeta}, \citenamefont
  {Blei}, \citenamefont {Carlstr\"{o}m}, \citenamefont {Watanabe},
  \citenamefont {Taniguchi}, \citenamefont {Tongay}, \citenamefont {Crommie},
  \citenamefont {Zettl},\ and\ \citenamefont {Wang}}]{Regan2020}%
  \BibitemOpen
  \bibfield  {author} {\bibinfo {author} {\bibfnamefont {E.~C.}\ \bibnamefont
  {Regan}}, \bibinfo {author} {\bibfnamefont {D.}~\bibnamefont {Wang}},
  \bibinfo {author} {\bibfnamefont {C.}~\bibnamefont {Jin}}, \bibinfo {author}
  {\bibfnamefont {M.~I.~B.}\ \bibnamefont {Utama}}, \bibinfo {author}
  {\bibfnamefont {B.}~\bibnamefont {Gao}}, \bibinfo {author} {\bibfnamefont
  {X.}~\bibnamefont {Wei}}, \bibinfo {author} {\bibfnamefont {S.}~\bibnamefont
  {Zhao}}, \bibinfo {author} {\bibfnamefont {W.}~\bibnamefont {Zhao}}, \bibinfo
  {author} {\bibfnamefont {Z.}~\bibnamefont {Zhang}}, \bibinfo {author}
  {\bibfnamefont {K.}~\bibnamefont {Yumigeta}}, \bibinfo {author}
  {\bibfnamefont {M.}~\bibnamefont {Blei}}, \bibinfo {author} {\bibfnamefont
  {J.~D.}\ \bibnamefont {Carlstr\"{o}m}}, \bibinfo {author} {\bibfnamefont
  {K.}~\bibnamefont {Watanabe}}, \bibinfo {author} {\bibfnamefont
  {T.}~\bibnamefont {Taniguchi}}, \bibinfo {author} {\bibfnamefont
  {S.}~\bibnamefont {Tongay}}, \bibinfo {author} {\bibfnamefont
  {M.}~\bibnamefont {Crommie}}, \bibinfo {author} {\bibfnamefont
  {A.}~\bibnamefont {Zettl}},\ and\ \bibinfo {author} {\bibfnamefont
  {F.}~\bibnamefont {Wang}},\ }\bibfield  {title} {\bibinfo {title} {Mott and
  generalized wigner crystal states in {WSe}2/{WS}2 moir{\'{e}}
  superlattices},\ }\href {https://doi.org/10.1038/s41586-020-2092-4}
  {\bibfield  {journal} {\bibinfo  {journal} {Nature}\ }\textbf {\bibinfo
  {volume} {579}},\ \bibinfo {pages} {359} (\bibinfo {year}
  {2020})}\BibitemShut {NoStop}%
\bibitem [{\citenamefont {Xian}\ \emph {et~al.}(2019)\citenamefont {Xian},
  \citenamefont {Kennes}, \citenamefont {Tancogne-Dejean}, \citenamefont
  {Altarelli},\ and\ \citenamefont {Rubio}}]{Xian2019}%
  \BibitemOpen
  \bibfield  {author} {\bibinfo {author} {\bibfnamefont {L.}~\bibnamefont
  {Xian}}, \bibinfo {author} {\bibfnamefont {D.~M.}\ \bibnamefont {Kennes}},
  \bibinfo {author} {\bibfnamefont {N.}~\bibnamefont {Tancogne-Dejean}},
  \bibinfo {author} {\bibfnamefont {M.}~\bibnamefont {Altarelli}},\ and\
  \bibinfo {author} {\bibfnamefont {A.}~\bibnamefont {Rubio}},\ }\bibfield
  {title} {\bibinfo {title} {Multiflat bands and strong correlations in twisted
  bilayer boron nitride: Doping-induced correlated insulator and
  superconductor},\ }\href {https://doi.org/10.1021/acs.nanolett.9b00986}
  {\bibfield  {journal} {\bibinfo  {journal} {Nano Letters}\ }\textbf {\bibinfo
  {volume} {19}},\ \bibinfo {pages} {4934} (\bibinfo {year}
  {2019})}\BibitemShut {NoStop}%
\bibitem [{\citenamefont {Kennes}\ \emph {et~al.}(2020)\citenamefont {Kennes},
  \citenamefont {Xian}, \citenamefont {Claassen},\ and\ \citenamefont
  {Rubio}}]{Kennes2020}%
  \BibitemOpen
  \bibfield  {author} {\bibinfo {author} {\bibfnamefont {D.~M.}\ \bibnamefont
  {Kennes}}, \bibinfo {author} {\bibfnamefont {L.}~\bibnamefont {Xian}},
  \bibinfo {author} {\bibfnamefont {M.}~\bibnamefont {Claassen}},\ and\
  \bibinfo {author} {\bibfnamefont {A.}~\bibnamefont {Rubio}},\ }\bibfield
  {title} {\bibinfo {title} {One-dimensional flat bands in twisted bilayer
  germanium selenide},\ }\href {https://doi.org/10.1038/s41467-020-14947-0}
  {\bibfield  {journal} {\bibinfo  {journal} {Nature Communications}\ }\textbf
  {\bibinfo {volume} {11}},\ \bibinfo {pages} {1124} (\bibinfo {year}
  {2020})}\BibitemShut {NoStop}%
\bibitem [{\citenamefont {Kennes}\ \emph {et~al.}(2021)\citenamefont {Kennes},
  \citenamefont {Claassen}, \citenamefont {Xian}, \citenamefont {Georges},
  \citenamefont {Millis}, \citenamefont {Hone}, \citenamefont {Dean},
  \citenamefont {Basov}, \citenamefont {Pasupathy},\ and\ \citenamefont
  {Rubio}}]{Kennes2021}%
  \BibitemOpen
  \bibfield  {author} {\bibinfo {author} {\bibfnamefont {D.~M.}\ \bibnamefont
  {Kennes}}, \bibinfo {author} {\bibfnamefont {M.}~\bibnamefont {Claassen}},
  \bibinfo {author} {\bibfnamefont {L.}~\bibnamefont {Xian}}, \bibinfo {author}
  {\bibfnamefont {A.}~\bibnamefont {Georges}}, \bibinfo {author} {\bibfnamefont
  {A.~J.}\ \bibnamefont {Millis}}, \bibinfo {author} {\bibfnamefont
  {J.}~\bibnamefont {Hone}}, \bibinfo {author} {\bibfnamefont {C.~R.}\
  \bibnamefont {Dean}}, \bibinfo {author} {\bibfnamefont {D.~N.}\ \bibnamefont
  {Basov}}, \bibinfo {author} {\bibfnamefont {A.~N.}\ \bibnamefont
  {Pasupathy}},\ and\ \bibinfo {author} {\bibfnamefont {A.}~\bibnamefont
  {Rubio}},\ }\bibfield  {title} {\bibinfo {title} {Moir{\'{e}}
  heterostructures as a condensed-matter quantum simulator},\ }\href
  {https://doi.org/10.1038/s41567-020-01154-3} {\bibfield  {journal} {\bibinfo
  {journal} {Nature Physics}\ }\textbf {\bibinfo {volume} {17}},\ \bibinfo
  {pages} {155} (\bibinfo {year} {2021})}\BibitemShut {NoStop}%
\bibitem [{\citenamefont {Carr}\ \emph {et~al.}(2020)\citenamefont {Carr},
  \citenamefont {Fang},\ and\ \citenamefont {Kaxiras}}]{Carr2020}%
  \BibitemOpen
  \bibfield  {author} {\bibinfo {author} {\bibfnamefont {S.}~\bibnamefont
  {Carr}}, \bibinfo {author} {\bibfnamefont {S.}~\bibnamefont {Fang}},\ and\
  \bibinfo {author} {\bibfnamefont {E.}~\bibnamefont {Kaxiras}},\ }\bibfield
  {title} {\bibinfo {title} {Electronic-structure methods for twisted
  moir{\'{e}} layers},\ }\href {https://doi.org/10.1038/s41578-020-0214-0}
  {\bibfield  {journal} {\bibinfo  {journal} {Nature Reviews Materials}\
  }\textbf {\bibinfo {volume} {5}},\ \bibinfo {pages} {748} (\bibinfo {year}
  {2020})}\BibitemShut {NoStop}%
\bibitem [{\citenamefont {Senior}\ \emph {et~al.}(2020)\citenamefont {Senior},
  \citenamefont {Evans}, \citenamefont {Jumper}, \citenamefont {Kirkpatrick},
  \citenamefont {Sifre}, \citenamefont {Green}, \citenamefont {Qin},
  \citenamefont {{\v{Z}}{\'{\i}}dek}, \citenamefont {Nelson}, \citenamefont
  {Bridgland}, \citenamefont {Penedones}, \citenamefont {Petersen},
  \citenamefont {Simonyan}, \citenamefont {Crossan}, \citenamefont {Kohli},
  \citenamefont {Jones}, \citenamefont {Silver}, \citenamefont {Kavukcuoglu},\
  and\ \citenamefont {Hassabis}}]{Senior2020}%
  \BibitemOpen
  \bibfield  {author} {\bibinfo {author} {\bibfnamefont {A.~W.}\ \bibnamefont
  {Senior}}, \bibinfo {author} {\bibfnamefont {R.}~\bibnamefont {Evans}},
  \bibinfo {author} {\bibfnamefont {J.}~\bibnamefont {Jumper}}, \bibinfo
  {author} {\bibfnamefont {J.}~\bibnamefont {Kirkpatrick}}, \bibinfo {author}
  {\bibfnamefont {L.}~\bibnamefont {Sifre}}, \bibinfo {author} {\bibfnamefont
  {T.}~\bibnamefont {Green}}, \bibinfo {author} {\bibfnamefont
  {C.}~\bibnamefont {Qin}}, \bibinfo {author} {\bibfnamefont {A.}~\bibnamefont
  {{\v{Z}}{\'{\i}}dek}}, \bibinfo {author} {\bibfnamefont {A.~W.~R.}\
  \bibnamefont {Nelson}}, \bibinfo {author} {\bibfnamefont {A.}~\bibnamefont
  {Bridgland}}, \bibinfo {author} {\bibfnamefont {H.}~\bibnamefont
  {Penedones}}, \bibinfo {author} {\bibfnamefont {S.}~\bibnamefont {Petersen}},
  \bibinfo {author} {\bibfnamefont {K.}~\bibnamefont {Simonyan}}, \bibinfo
  {author} {\bibfnamefont {S.}~\bibnamefont {Crossan}}, \bibinfo {author}
  {\bibfnamefont {P.}~\bibnamefont {Kohli}}, \bibinfo {author} {\bibfnamefont
  {D.~T.}\ \bibnamefont {Jones}}, \bibinfo {author} {\bibfnamefont
  {D.}~\bibnamefont {Silver}}, \bibinfo {author} {\bibfnamefont
  {K.}~\bibnamefont {Kavukcuoglu}},\ and\ \bibinfo {author} {\bibfnamefont
  {D.}~\bibnamefont {Hassabis}},\ }\bibfield  {title} {\bibinfo {title}
  {Improved protein structure prediction using potentials from deep learning},\
  }\href {https://doi.org/10.1038/s41586-019-1923-7} {\bibfield  {journal}
  {\bibinfo  {journal} {Nature}\ }\textbf {\bibinfo {volume} {577}},\ \bibinfo
  {pages} {706} (\bibinfo {year} {2020})}\BibitemShut {NoStop}%
\bibitem [{\citenamefont {Zhang}\ \emph {et~al.}(2018)\citenamefont {Zhang},
  \citenamefont {Han}, \citenamefont {Wang}, \citenamefont {Car},\ and\
  \citenamefont {E}}]{Zhang2018}%
  \BibitemOpen
  \bibfield  {author} {\bibinfo {author} {\bibfnamefont {L.}~\bibnamefont
  {Zhang}}, \bibinfo {author} {\bibfnamefont {J.}~\bibnamefont {Han}}, \bibinfo
  {author} {\bibfnamefont {H.}~\bibnamefont {Wang}}, \bibinfo {author}
  {\bibfnamefont {R.}~\bibnamefont {Car}},\ and\ \bibinfo {author}
  {\bibfnamefont {W.}~\bibnamefont {E}},\ }\bibfield  {title} {\bibinfo {title}
  {Deep potential molecular dynamics: A scalable model with the accuracy of
  quantum mechanics},\ }\href {https://doi.org/10.1103/physrevlett.120.143001}
  {\bibfield  {journal} {\bibinfo  {journal} {Physical Review Letters}\
  }\textbf {\bibinfo {volume} {120}},\ \bibinfo {pages} {143001} (\bibinfo
  {year} {2018})}\BibitemShut {NoStop}%
\bibitem [{\citenamefont {Behler}\ and\ \citenamefont
  {Parrinello}(2007)}]{behler2007generalized}%
  \BibitemOpen
  \bibfield  {author} {\bibinfo {author} {\bibfnamefont {J.}~\bibnamefont
  {Behler}}\ and\ \bibinfo {author} {\bibfnamefont {M.}~\bibnamefont
  {Parrinello}},\ }\bibfield  {title} {\bibinfo {title} {Generalized
  neural-network representation of high-dimensional potential-energy
  surfaces},\ }\href@noop {} {\bibfield  {journal} {\bibinfo  {journal}
  {Physical review letters}\ }\textbf {\bibinfo {volume} {98}},\ \bibinfo
  {pages} {146401} (\bibinfo {year} {2007})}\BibitemShut {NoStop}%
\bibitem [{\citenamefont {Li}\ \emph {et~al.}(2022)\citenamefont {Li},
  \citenamefont {Wang}, \citenamefont {Zou}, \citenamefont {Ye}, \citenamefont
  {Xu}, \citenamefont {Gong}, \citenamefont {Duan},\ and\ \citenamefont
  {Xu}}]{li_deep-learning_2022}%
  \BibitemOpen
  \bibfield  {author} {\bibinfo {author} {\bibfnamefont {H.}~\bibnamefont
  {Li}}, \bibinfo {author} {\bibfnamefont {Z.}~\bibnamefont {Wang}}, \bibinfo
  {author} {\bibfnamefont {N.}~\bibnamefont {Zou}}, \bibinfo {author}
  {\bibfnamefont {M.}~\bibnamefont {Ye}}, \bibinfo {author} {\bibfnamefont
  {R.}~\bibnamefont {Xu}}, \bibinfo {author} {\bibfnamefont {X.}~\bibnamefont
  {Gong}}, \bibinfo {author} {\bibfnamefont {W.}~\bibnamefont {Duan}},\ and\
  \bibinfo {author} {\bibfnamefont {Y.}~\bibnamefont {Xu}},\ }\bibfield
  {title} {\bibinfo {title} {Deep-learning density functional theory
  {Hamiltonian} for efficient ab initio electronic-structure calculation},\
  }\href {https://doi.org/10.1038/s43588-022-00265-6} {\bibfield  {journal}
  {\bibinfo  {journal} {Nature Computational Science}\ }\textbf {\bibinfo
  {volume} {2}},\ \bibinfo {pages} {367} (\bibinfo {year} {2022})}\BibitemShut
  {NoStop}%
\bibitem [{\citenamefont {Gong}\ \emph {et~al.}(2023)\citenamefont {Gong},
  \citenamefont {Li}, \citenamefont {Zou}, \citenamefont {Xu}, \citenamefont
  {Duan},\ and\ \citenamefont {Xu}}]{gong2023general}%
  \BibitemOpen
  \bibfield  {author} {\bibinfo {author} {\bibfnamefont {X.}~\bibnamefont
  {Gong}}, \bibinfo {author} {\bibfnamefont {H.}~\bibnamefont {Li}}, \bibinfo
  {author} {\bibfnamefont {N.}~\bibnamefont {Zou}}, \bibinfo {author}
  {\bibfnamefont {R.}~\bibnamefont {Xu}}, \bibinfo {author} {\bibfnamefont
  {W.}~\bibnamefont {Duan}},\ and\ \bibinfo {author} {\bibfnamefont
  {Y.}~\bibnamefont {Xu}},\ }\bibfield  {title} {\bibinfo {title} {General
  framework for {{E}}(3)-equivariant neural network representation of density
  functional theory {{Hamiltonian}}},\ }\href
  {https://doi.org/10.1038/s41467-023-38468-8} {\bibfield  {journal} {\bibinfo
  {journal} {Nature Communications}\ }\textbf {\bibinfo {volume} {14}},\
  \bibinfo {pages} {2848} (\bibinfo {year} {2023})}\BibitemShut {NoStop}%
\bibitem [{\citenamefont {Jain}\ \emph {et~al.}(2013)\citenamefont {Jain},
  \citenamefont {Ong}, \citenamefont {Hautier}, \citenamefont {Chen},
  \citenamefont {Richards}, \citenamefont {Dacek}, \citenamefont {Cholia},
  \citenamefont {Gunter}, \citenamefont {Skinner}, \citenamefont {Ceder},\ and\
  \citenamefont {Persson}}]{Jain2013}%
  \BibitemOpen
  \bibfield  {author} {\bibinfo {author} {\bibfnamefont {A.}~\bibnamefont
  {Jain}}, \bibinfo {author} {\bibfnamefont {S.~P.}\ \bibnamefont {Ong}},
  \bibinfo {author} {\bibfnamefont {G.}~\bibnamefont {Hautier}}, \bibinfo
  {author} {\bibfnamefont {W.}~\bibnamefont {Chen}}, \bibinfo {author}
  {\bibfnamefont {W.~D.}\ \bibnamefont {Richards}}, \bibinfo {author}
  {\bibfnamefont {S.}~\bibnamefont {Dacek}}, \bibinfo {author} {\bibfnamefont
  {S.}~\bibnamefont {Cholia}}, \bibinfo {author} {\bibfnamefont
  {D.}~\bibnamefont {Gunter}}, \bibinfo {author} {\bibfnamefont
  {D.}~\bibnamefont {Skinner}}, \bibinfo {author} {\bibfnamefont
  {G.}~\bibnamefont {Ceder}},\ and\ \bibinfo {author} {\bibfnamefont {K.~A.}\
  \bibnamefont {Persson}},\ }\bibfield  {title} {\bibinfo {title} {Commentary:
  The materials project: A materials genome approach to accelerating materials
  innovation},\ }\href {https://doi.org/10.1063/1.4812323} {\bibfield
  {journal} {\bibinfo  {journal} {{APL} Materials}\ }\textbf {\bibinfo {volume}
  {1}},\ \bibinfo {pages} {011002} (\bibinfo {year} {2013})}\BibitemShut
  {NoStop}%
\bibitem [{\citenamefont {Hellenbrandt}(2004)}]{hellenbrandt2004inorganic}%
  \BibitemOpen
  \bibfield  {author} {\bibinfo {author} {\bibfnamefont {M.}~\bibnamefont
  {Hellenbrandt}},\ }\bibfield  {title} {\bibinfo {title} {The inorganic
  crystal structure database (icsd)—present and future},\ }\href@noop {}
  {\bibfield  {journal} {\bibinfo  {journal} {Crystallography Reviews}\
  }\textbf {\bibinfo {volume} {10}},\ \bibinfo {pages} {17} (\bibinfo {year}
  {2004})}\BibitemShut {NoStop}%
\bibitem [{\citenamefont {Haastrup}\ \emph {et~al.}(2018)\citenamefont
  {Haastrup}, \citenamefont {Strange}, \citenamefont {Pandey}, \citenamefont
  {Deilmann}, \citenamefont {Schmidt}, \citenamefont {Hinsche}, \citenamefont
  {Gjerding}, \citenamefont {Torelli}, \citenamefont {Larsen}, \citenamefont
  {{Riis-Jensen}}, \citenamefont {Gath}, \citenamefont {Jacobsen},
  \citenamefont {Mortensen}, \citenamefont {Olsen},\ and\ \citenamefont
  {Thygesen}}]{haastrupComputational2018}%
  \BibitemOpen
  \bibfield  {author} {\bibinfo {author} {\bibfnamefont {S.}~\bibnamefont
  {Haastrup}}, \bibinfo {author} {\bibfnamefont {M.}~\bibnamefont {Strange}},
  \bibinfo {author} {\bibfnamefont {M.}~\bibnamefont {Pandey}}, \bibinfo
  {author} {\bibfnamefont {T.}~\bibnamefont {Deilmann}}, \bibinfo {author}
  {\bibfnamefont {P.~S.}\ \bibnamefont {Schmidt}}, \bibinfo {author}
  {\bibfnamefont {N.~F.}\ \bibnamefont {Hinsche}}, \bibinfo {author}
  {\bibfnamefont {M.~N.}\ \bibnamefont {Gjerding}}, \bibinfo {author}
  {\bibfnamefont {D.}~\bibnamefont {Torelli}}, \bibinfo {author} {\bibfnamefont
  {P.~M.}\ \bibnamefont {Larsen}}, \bibinfo {author} {\bibfnamefont {A.~C.}\
  \bibnamefont {{Riis-Jensen}}}, \bibinfo {author} {\bibfnamefont
  {J.}~\bibnamefont {Gath}}, \bibinfo {author} {\bibfnamefont {K.~W.}\
  \bibnamefont {Jacobsen}}, \bibinfo {author} {\bibfnamefont {J.~J.}\
  \bibnamefont {Mortensen}}, \bibinfo {author} {\bibfnamefont {T.}~\bibnamefont
  {Olsen}},\ and\ \bibinfo {author} {\bibfnamefont {K.~S.}\ \bibnamefont
  {Thygesen}},\ }\bibfield  {title} {\bibinfo {title} {The {{Computational 2D
  Materials Database}}: High-throughput modeling and discovery of atomically
  thin crystals},\ }\href {https://doi.org/10.1088/2053-1583/aacfc1} {\bibfield
   {journal} {\bibinfo  {journal} {2D Materials}\ }\textbf {\bibinfo {volume}
  {5}},\ \bibinfo {pages} {042002} (\bibinfo {year} {2018})}\BibitemShut
  {NoStop}%
\bibitem [{\citenamefont {Gjerding}\ \emph {et~al.}(2021)\citenamefont
  {Gjerding}, \citenamefont {Taghizadeh}, \citenamefont {Rasmussen},
  \citenamefont {Ali}, \citenamefont {Bertoldo}, \citenamefont {Deilmann},
  \citenamefont {Kn{\o}sgaard}, \citenamefont {Kruse}, \citenamefont {Larsen},
  \citenamefont {Manti}, \citenamefont {Pedersen}, \citenamefont {Petralanda},
  \citenamefont {Skovhus}, \citenamefont {Svendsen}, \citenamefont {Mortensen},
  \citenamefont {Olsen},\ and\ \citenamefont {Thygesen}}]{gjerdingRecent2021}%
  \BibitemOpen
  \bibfield  {author} {\bibinfo {author} {\bibfnamefont {M.~N.}\ \bibnamefont
  {Gjerding}}, \bibinfo {author} {\bibfnamefont {A.}~\bibnamefont
  {Taghizadeh}}, \bibinfo {author} {\bibfnamefont {A.}~\bibnamefont
  {Rasmussen}}, \bibinfo {author} {\bibfnamefont {S.}~\bibnamefont {Ali}},
  \bibinfo {author} {\bibfnamefont {F.}~\bibnamefont {Bertoldo}}, \bibinfo
  {author} {\bibfnamefont {T.}~\bibnamefont {Deilmann}}, \bibinfo {author}
  {\bibfnamefont {N.~R.}\ \bibnamefont {Kn{\o}sgaard}}, \bibinfo {author}
  {\bibfnamefont {M.}~\bibnamefont {Kruse}}, \bibinfo {author} {\bibfnamefont
  {A.~H.}\ \bibnamefont {Larsen}}, \bibinfo {author} {\bibfnamefont
  {S.}~\bibnamefont {Manti}}, \bibinfo {author} {\bibfnamefont {T.~G.}\
  \bibnamefont {Pedersen}}, \bibinfo {author} {\bibfnamefont {U.}~\bibnamefont
  {Petralanda}}, \bibinfo {author} {\bibfnamefont {T.}~\bibnamefont {Skovhus}},
  \bibinfo {author} {\bibfnamefont {M.~K.}\ \bibnamefont {Svendsen}}, \bibinfo
  {author} {\bibfnamefont {J.~J.}\ \bibnamefont {Mortensen}}, \bibinfo {author}
  {\bibfnamefont {T.}~\bibnamefont {Olsen}},\ and\ \bibinfo {author}
  {\bibfnamefont {K.~S.}\ \bibnamefont {Thygesen}},\ }\bibfield  {title}
  {\bibinfo {title} {Recent progress of the {{Computational 2D Materials
  Database}} ({{C2DB}})},\ }\href {https://doi.org/10.1088/2053-1583/ac1059}
  {\bibfield  {journal} {\bibinfo  {journal} {2D Materials}\ }\textbf {\bibinfo
  {volume} {8}},\ \bibinfo {pages} {044002} (\bibinfo {year}
  {2021})}\BibitemShut {NoStop}%
\bibitem [{\citenamefont {Jacobs}\ \emph {et~al.}(1991)\citenamefont {Jacobs},
  \citenamefont {Jordan}, \citenamefont {Nowlan},\ and\ \citenamefont
  {Hinton}}]{jacobs1991adaptive}%
  \BibitemOpen
  \bibfield  {author} {\bibinfo {author} {\bibfnamefont {R.~A.}\ \bibnamefont
  {Jacobs}}, \bibinfo {author} {\bibfnamefont {M.~I.}\ \bibnamefont {Jordan}},
  \bibinfo {author} {\bibfnamefont {S.~J.}\ \bibnamefont {Nowlan}},\ and\
  \bibinfo {author} {\bibfnamefont {G.~E.}\ \bibnamefont {Hinton}},\ }\bibfield
   {title} {\bibinfo {title} {Adaptive mixtures of local experts},\ }\href@noop
  {} {\bibfield  {journal} {\bibinfo  {journal} {Neural computation}\ }\textbf
  {\bibinfo {volume} {3}},\ \bibinfo {pages} {79} (\bibinfo {year}
  {1991})}\BibitemShut {NoStop}%
\bibitem [{\citenamefont {Kresse}\ and\ \citenamefont
  {Furthm{\"u}ller}(1996{\natexlab{a}})}]{kresse1996_1}%
  \BibitemOpen
  \bibfield  {author} {\bibinfo {author} {\bibfnamefont {G.}~\bibnamefont
  {Kresse}}\ and\ \bibinfo {author} {\bibfnamefont {J.}~\bibnamefont
  {Furthm{\"u}ller}},\ }\bibfield  {title} {\bibinfo {title} {Efficiency of
  ab-initio total energy calculations for metals and semiconductors using a
  plane-wave basis set},\ }\href {https://doi.org/10.1016/0927-0256(96)00008-0}
  {\bibfield  {journal} {\bibinfo  {journal} {Computational Materials Science}\
  }\textbf {\bibinfo {volume} {6}},\ \bibinfo {pages} {15} (\bibinfo {year}
  {1996}{\natexlab{a}})}\BibitemShut {NoStop}%
\bibitem [{\citenamefont {Kresse}\ and\ \citenamefont
  {Furthm{\"u}ller}(1996{\natexlab{b}})}]{kresse1996efficient}%
  \BibitemOpen
  \bibfield  {author} {\bibinfo {author} {\bibfnamefont {G.}~\bibnamefont
  {Kresse}}\ and\ \bibinfo {author} {\bibfnamefont {J.}~\bibnamefont
  {Furthm{\"u}ller}},\ }\bibfield  {title} {\bibinfo {title} {Efficient
  iterative schemes for ab initio total-energy calculations using a plane-wave
  basis set},\ }\href@noop {} {\bibfield  {journal} {\bibinfo  {journal}
  {Physical review B}\ }\textbf {\bibinfo {volume} {54}},\ \bibinfo {pages}
  {11169} (\bibinfo {year} {1996}{\natexlab{b}})}\BibitemShut {NoStop}%
\bibitem [{\citenamefont {Xie}\ and\ \citenamefont
  {Grossman}(2018)}]{PhysRevLett.120.145301}%
  \BibitemOpen
  \bibfield  {author} {\bibinfo {author} {\bibfnamefont {T.}~\bibnamefont
  {Xie}}\ and\ \bibinfo {author} {\bibfnamefont {J.~C.}\ \bibnamefont
  {Grossman}},\ }\bibfield  {title} {\bibinfo {title} {Crystal graph
  convolutional neural networks for an accurate and interpretable prediction of
  material properties},\ }\href
  {https://doi.org/10.1103/PhysRevLett.120.145301} {\bibfield  {journal}
  {\bibinfo  {journal} {Physical Review Letters}\ }\textbf {\bibinfo {volume}
  {120}},\ \bibinfo {pages} {145301} (\bibinfo {year} {2018})}\BibitemShut
  {NoStop}%
\bibitem [{\citenamefont {Manzeli}\ \emph {et~al.}(2017)\citenamefont
  {Manzeli}, \citenamefont {Ovchinnikov}, \citenamefont {Pasquier},
  \citenamefont {Yazyev},\ and\ \citenamefont {Kis}}]{Manzeli2017}%
  \BibitemOpen
  \bibfield  {author} {\bibinfo {author} {\bibfnamefont {S.}~\bibnamefont
  {Manzeli}}, \bibinfo {author} {\bibfnamefont {D.}~\bibnamefont
  {Ovchinnikov}}, \bibinfo {author} {\bibfnamefont {D.}~\bibnamefont
  {Pasquier}}, \bibinfo {author} {\bibfnamefont {O.~V.}\ \bibnamefont
  {Yazyev}},\ and\ \bibinfo {author} {\bibfnamefont {A.}~\bibnamefont {Kis}},\
  }\bibfield  {title} {\bibinfo {title} {2d transition metal dichalcogenides},\
  }\href {https://doi.org/10.1038/natrevmats.2017.33} {\bibfield  {journal}
  {\bibinfo  {journal} {Nature Reviews Materials}\ }\textbf {\bibinfo {volume}
  {2}},\ \bibinfo {pages} {17033} (\bibinfo {year} {2017})}\BibitemShut
  {NoStop}%
\bibitem [{\citenamefont {Deringer}\ \emph {et~al.}(2020)\citenamefont
  {Deringer}, \citenamefont {Caro},\ and\ \citenamefont
  {Cs{\'a}nyi}}]{deringer2020general}%
  \BibitemOpen
  \bibfield  {author} {\bibinfo {author} {\bibfnamefont {V.~L.}\ \bibnamefont
  {Deringer}}, \bibinfo {author} {\bibfnamefont {M.~A.}\ \bibnamefont {Caro}},\
  and\ \bibinfo {author} {\bibfnamefont {G.}~\bibnamefont {Cs{\'a}nyi}},\
  }\bibfield  {title} {\bibinfo {title} {A general-purpose machine-learning
  force field for bulk and nanostructured phosphorus},\ }\href
  {https://doi.org/10.1038/s41467-020-19168-z} {\bibfield  {journal} {\bibinfo
  {journal} {Nature Communications}\ }\textbf {\bibinfo {volume} {11}},\
  \bibinfo {pages} {5461} (\bibinfo {year} {2020})}\BibitemShut {NoStop}%
\bibitem [{\citenamefont {Li}\ \emph {et~al.}(2023)\citenamefont {Li},
  \citenamefont {Tang}, \citenamefont {Gong}, \citenamefont {Zou},
  \citenamefont {Duan},\ and\ \citenamefont {Xu}}]{xdeeph}%
  \BibitemOpen
  \bibfield  {author} {\bibinfo {author} {\bibfnamefont {H.}~\bibnamefont
  {Li}}, \bibinfo {author} {\bibfnamefont {Z.}~\bibnamefont {Tang}}, \bibinfo
  {author} {\bibfnamefont {X.}~\bibnamefont {Gong}}, \bibinfo {author}
  {\bibfnamefont {N.}~\bibnamefont {Zou}}, \bibinfo {author} {\bibfnamefont
  {W.}~\bibnamefont {Duan}},\ and\ \bibinfo {author} {\bibfnamefont
  {Y.}~\bibnamefont {Xu}},\ }\bibfield  {title} {\bibinfo {title}
  {Deep-learning electronic-structure calculation of magnetic
  superstructures},\ }\href {https://doi.org/10.1038/s43588-023-00424-3}
  {\bibfield  {journal} {\bibinfo  {journal} {Nature Computational Science}\
  }\textbf {\bibinfo {volume} {3}},\ \bibinfo {pages} {321} (\bibinfo {year}
  {2023})}\BibitemShut {NoStop}%
\bibitem [{\citenamefont {Bradlyn}\ \emph {et~al.}(2017)\citenamefont
  {Bradlyn}, \citenamefont {Elcoro}, \citenamefont {Cano}, \citenamefont
  {Vergniory}, \citenamefont {Wang}, \citenamefont {Felser}, \citenamefont
  {Aroyo},\ and\ \citenamefont {Bernevig}}]{Bradlyn2017}%
  \BibitemOpen
  \bibfield  {author} {\bibinfo {author} {\bibfnamefont {B.}~\bibnamefont
  {Bradlyn}}, \bibinfo {author} {\bibfnamefont {L.}~\bibnamefont {Elcoro}},
  \bibinfo {author} {\bibfnamefont {J.}~\bibnamefont {Cano}}, \bibinfo {author}
  {\bibfnamefont {M.~G.}\ \bibnamefont {Vergniory}}, \bibinfo {author}
  {\bibfnamefont {Z.}~\bibnamefont {Wang}}, \bibinfo {author} {\bibfnamefont
  {C.}~\bibnamefont {Felser}}, \bibinfo {author} {\bibfnamefont {M.~I.}\
  \bibnamefont {Aroyo}},\ and\ \bibinfo {author} {\bibfnamefont {B.~A.}\
  \bibnamefont {Bernevig}},\ }\bibfield  {title} {\bibinfo {title} {Topological
  quantum chemistry},\ }\href {https://doi.org/10.1038/nature23268} {\bibfield
  {journal} {\bibinfo  {journal} {Nature}\ }\textbf {\bibinfo {volume} {547}},\
  \bibinfo {pages} {298} (\bibinfo {year} {2017})}\BibitemShut {NoStop}%
\bibitem [{\citenamefont {Vergniory}\ \emph {et~al.}(2019)\citenamefont
  {Vergniory}, \citenamefont {Elcoro}, \citenamefont {Felser}, \citenamefont
  {Regnault}, \citenamefont {Bernevig},\ and\ \citenamefont
  {Wang}}]{Vergniory2019}%
  \BibitemOpen
  \bibfield  {author} {\bibinfo {author} {\bibfnamefont {M.~G.}\ \bibnamefont
  {Vergniory}}, \bibinfo {author} {\bibfnamefont {L.}~\bibnamefont {Elcoro}},
  \bibinfo {author} {\bibfnamefont {C.}~\bibnamefont {Felser}}, \bibinfo
  {author} {\bibfnamefont {N.}~\bibnamefont {Regnault}}, \bibinfo {author}
  {\bibfnamefont {B.~A.}\ \bibnamefont {Bernevig}},\ and\ \bibinfo {author}
  {\bibfnamefont {Z.}~\bibnamefont {Wang}},\ }\bibfield  {title} {\bibinfo
  {title} {A complete catalogue of high-quality topological materials},\ }\href
  {https://doi.org/10.1038/s41586-019-0954-4} {\bibfield  {journal} {\bibinfo
  {journal} {Nature}\ }\textbf {\bibinfo {volume} {566}},\ \bibinfo {pages}
  {480} (\bibinfo {year} {2019})}\BibitemShut {NoStop}%
\bibitem [{\citenamefont {Vergniory}\ \emph {et~al.}(2022)\citenamefont
  {Vergniory}, \citenamefont {Wieder}, \citenamefont {Elcoro}, \citenamefont
  {Parkin}, \citenamefont {Felser}, \citenamefont {Bernevig},\ and\
  \citenamefont {Regnault}}]{Vergniory2022}%
  \BibitemOpen
  \bibfield  {author} {\bibinfo {author} {\bibfnamefont {M.~G.}\ \bibnamefont
  {Vergniory}}, \bibinfo {author} {\bibfnamefont {B.~J.}\ \bibnamefont
  {Wieder}}, \bibinfo {author} {\bibfnamefont {L.}~\bibnamefont {Elcoro}},
  \bibinfo {author} {\bibfnamefont {S.~S.~P.}\ \bibnamefont {Parkin}}, \bibinfo
  {author} {\bibfnamefont {C.}~\bibnamefont {Felser}}, \bibinfo {author}
  {\bibfnamefont {B.~A.}\ \bibnamefont {Bernevig}},\ and\ \bibinfo {author}
  {\bibfnamefont {N.}~\bibnamefont {Regnault}},\ }\bibfield  {title} {\bibinfo
  {title} {All topological bands of all nonmagnetic stoichiometric materials},\
  }\href {https://doi.org/10.1126/science.abg9094} {\bibfield  {journal}
  {\bibinfo  {journal} {Science}\ }\textbf {\bibinfo {volume} {376}},\ \bibinfo
  {pages} {eabg9094} (\bibinfo {year} {2022})}\BibitemShut {NoStop}%
\bibitem [{\citenamefont {C{\u{a}}lug{\u{a}}ru}\ \emph
  {et~al.}(2021)\citenamefont {C{\u{a}}lug{\u{a}}ru}, \citenamefont {Chew},
  \citenamefont {Elcoro}, \citenamefont {Xu}, \citenamefont {Regnault},
  \citenamefont {Song},\ and\ \citenamefont {Bernevig}}]{Clugru2021}%
  \BibitemOpen
  \bibfield  {author} {\bibinfo {author} {\bibfnamefont {D.}~\bibnamefont
  {C{\u{a}}lug{\u{a}}ru}}, \bibinfo {author} {\bibfnamefont {A.}~\bibnamefont
  {Chew}}, \bibinfo {author} {\bibfnamefont {L.}~\bibnamefont {Elcoro}},
  \bibinfo {author} {\bibfnamefont {Y.}~\bibnamefont {Xu}}, \bibinfo {author}
  {\bibfnamefont {N.}~\bibnamefont {Regnault}}, \bibinfo {author}
  {\bibfnamefont {Z.-D.}\ \bibnamefont {Song}},\ and\ \bibinfo {author}
  {\bibfnamefont {B.~A.}\ \bibnamefont {Bernevig}},\ }\bibfield  {title}
  {\bibinfo {title} {General construction and topological classification of
  crystalline flat bands},\ }\href {https://doi.org/10.1038/s41567-021-01445-3}
  {\bibfield  {journal} {\bibinfo  {journal} {Nature Physics}\ }\textbf
  {\bibinfo {volume} {18}},\ \bibinfo {pages} {185} (\bibinfo {year}
  {2021})}\BibitemShut {NoStop}%
\bibitem [{\citenamefont {Kresse}\ and\ \citenamefont
  {Furthm\"uller}(1996)}]{PhysRevB.54.11169}%
  \BibitemOpen
  \bibfield  {author} {\bibinfo {author} {\bibfnamefont {G.}~\bibnamefont
  {Kresse}}\ and\ \bibinfo {author} {\bibfnamefont {J.}~\bibnamefont
  {Furthm\"uller}},\ }\bibfield  {title} {\bibinfo {title} {Efficient iterative
  schemes for ab initio total-energy calculations using a plane-wave basis
  set},\ }\href {https://doi.org/10.1103/PhysRevB.54.11169} {\bibfield
  {journal} {\bibinfo  {journal} {Physical Review B}\ }\textbf {\bibinfo
  {volume} {54}},\ \bibinfo {pages} {11169} (\bibinfo {year}
  {1996})}\BibitemShut {NoStop}%
\bibitem [{\citenamefont {Bl\"ochl}(1994)}]{PhysRevB.50.17953}%
  \BibitemOpen
  \bibfield  {author} {\bibinfo {author} {\bibfnamefont {P.~E.}\ \bibnamefont
  {Bl\"ochl}},\ }\bibfield  {title} {\bibinfo {title} {Projector augmented-wave
  method},\ }\href {https://doi.org/10.1103/PhysRevB.50.17953} {\bibfield
  {journal} {\bibinfo  {journal} {Physical Review B}\ }\textbf {\bibinfo
  {volume} {50}},\ \bibinfo {pages} {17953} (\bibinfo {year}
  {1994})}\BibitemShut {NoStop}%
\bibitem [{\citenamefont {Kresse}\ and\ \citenamefont
  {Joubert}(1999)}]{PhysRevB.59.1758}%
  \BibitemOpen
  \bibfield  {author} {\bibinfo {author} {\bibfnamefont {G.}~\bibnamefont
  {Kresse}}\ and\ \bibinfo {author} {\bibfnamefont {D.}~\bibnamefont
  {Joubert}},\ }\bibfield  {title} {\bibinfo {title} {From ultrasoft
  pseudopotentials to the projector augmented-wave method},\ }\href
  {https://doi.org/10.1103/PhysRevB.59.1758} {\bibfield  {journal} {\bibinfo
  {journal} {Physical Review B}\ }\textbf {\bibinfo {volume} {59}},\ \bibinfo
  {pages} {1758} (\bibinfo {year} {1999})}\BibitemShut {NoStop}%
\bibitem [{\citenamefont {Perdew}\ \emph {et~al.}(1996)\citenamefont {Perdew},
  \citenamefont {Burke},\ and\ \citenamefont
  {Ernzerhof}}]{PhysRevLett.77.3865}%
  \BibitemOpen
  \bibfield  {author} {\bibinfo {author} {\bibfnamefont {J.~P.}\ \bibnamefont
  {Perdew}}, \bibinfo {author} {\bibfnamefont {K.}~\bibnamefont {Burke}},\ and\
  \bibinfo {author} {\bibfnamefont {M.}~\bibnamefont {Ernzerhof}},\ }\bibfield
  {title} {\bibinfo {title} {Generalized gradient approximation made simple},\
  }\href {https://doi.org/10.1103/PhysRevLett.77.3865} {\bibfield  {journal}
  {\bibinfo  {journal} {Physical Review Letters}\ }\textbf {\bibinfo {volume}
  {77}},\ \bibinfo {pages} {3865} (\bibinfo {year} {1996})}\BibitemShut
  {NoStop}%
\bibitem [{\citenamefont {Grimme}\ \emph {et~al.}(2011)\citenamefont {Grimme},
  \citenamefont {Ehrlich},\ and\ \citenamefont {Goerigk}}]{RN1}%
  \BibitemOpen
  \bibfield  {author} {\bibinfo {author} {\bibfnamefont {S.}~\bibnamefont
  {Grimme}}, \bibinfo {author} {\bibfnamefont {S.}~\bibnamefont {Ehrlich}},\
  and\ \bibinfo {author} {\bibfnamefont {L.}~\bibnamefont {Goerigk}},\
  }\bibfield  {title} {\bibinfo {title} {Effect of the damping function in
  dispersion corrected density functional theory},\ }\href
  {https://doi.org/https://doi.org/10.1002/jcc.21759} {\bibfield  {journal}
  {\bibinfo  {journal} {Journal of Computational Chemistry}\ }\textbf {\bibinfo
  {volume} {32}},\ \bibinfo {pages} {1456} (\bibinfo {year}
  {2011})}\BibitemShut {NoStop}%
\bibitem [{\citenamefont {Grimme}\ \emph {et~al.}(2010)\citenamefont {Grimme},
  \citenamefont {Antony}, \citenamefont {Ehrlich},\ and\ \citenamefont
  {Krieg}}]{RN2}%
  \BibitemOpen
  \bibfield  {author} {\bibinfo {author} {\bibfnamefont {S.}~\bibnamefont
  {Grimme}}, \bibinfo {author} {\bibfnamefont {J.}~\bibnamefont {Antony}},
  \bibinfo {author} {\bibfnamefont {S.}~\bibnamefont {Ehrlich}},\ and\ \bibinfo
  {author} {\bibfnamefont {H.}~\bibnamefont {Krieg}},\ }\bibfield  {title}
  {\bibinfo {title} {A consistent and accurate ab initio parametrization of
  density functional dispersion correction (dft-d) for the 94 elements h-pu},\
  }\href {https://doi.org/10.1063/1.3382344} {\bibfield  {journal} {\bibinfo
  {journal} {The Journal of Chemical Physics}\ }\textbf {\bibinfo {volume}
  {132}},\ \bibinfo {pages} {154104} (\bibinfo {year} {2010})}\BibitemShut
  {NoStop}%
\bibitem [{\citenamefont {Monkhorst}\ and\ \citenamefont
  {Pack}(1976)}]{PhysRevB.13.5188}%
  \BibitemOpen
  \bibfield  {author} {\bibinfo {author} {\bibfnamefont {H.~J.}\ \bibnamefont
  {Monkhorst}}\ and\ \bibinfo {author} {\bibfnamefont {J.~D.}\ \bibnamefont
  {Pack}},\ }\bibfield  {title} {\bibinfo {title} {Special points for
  brillouin-zone integrations},\ }\href
  {https://doi.org/10.1103/PhysRevB.13.5188} {\bibfield  {journal} {\bibinfo
  {journal} {Physical Review B}\ }\textbf {\bibinfo {volume} {13}},\ \bibinfo
  {pages} {5188} (\bibinfo {year} {1976})}\BibitemShut {NoStop}%
\bibitem [{\citenamefont {Ozaki}(2003)}]{Ozaki2003}%
  \BibitemOpen
  \bibfield  {author} {\bibinfo {author} {\bibfnamefont {T.}~\bibnamefont
  {Ozaki}},\ }\bibfield  {title} {\bibinfo {title} {Variationally optimized
  atomic orbitals for large-scale electronic structures},\ }\href
  {https://doi.org/10.1103/PhysRevB.67.155108} {\bibfield  {journal} {\bibinfo
  {journal} {Phys. Rev. B}\ }\textbf {\bibinfo {volume} {67}},\ \bibinfo
  {pages} {155108} (\bibinfo {year} {2003})}\BibitemShut {NoStop}%
\bibitem [{\citenamefont {Ozaki}\ and\ \citenamefont {Kino}(2004)}]{Ozaki2004}%
  \BibitemOpen
  \bibfield  {author} {\bibinfo {author} {\bibfnamefont {T.}~\bibnamefont
  {Ozaki}}\ and\ \bibinfo {author} {\bibfnamefont {H.}~\bibnamefont {Kino}},\
  }\bibfield  {title} {\bibinfo {title} {Numerical atomic basis orbitals from h
  to kr},\ }\href {https://doi.org/10.1103/PhysRevB.69.195113} {\bibfield
  {journal} {\bibinfo  {journal} {Phys. Rev. B}\ }\textbf {\bibinfo {volume}
  {69}},\ \bibinfo {pages} {195113} (\bibinfo {year} {2004})}\BibitemShut
  {NoStop}%
\bibitem [{\citenamefont {Morrison}\ \emph {et~al.}(1993)\citenamefont
  {Morrison}, \citenamefont {Bylander},\ and\ \citenamefont
  {Kleinman}}]{Morrison1993}%
  \BibitemOpen
  \bibfield  {author} {\bibinfo {author} {\bibfnamefont {I.}~\bibnamefont
  {Morrison}}, \bibinfo {author} {\bibfnamefont {D.~M.}\ \bibnamefont
  {Bylander}},\ and\ \bibinfo {author} {\bibfnamefont {L.}~\bibnamefont
  {Kleinman}},\ }\bibfield  {title} {\bibinfo {title} {Nonlocal hermitian
  norm-conserving vanderbilt pseudopotential},\ }\href
  {https://doi.org/10.1103/PhysRevB.47.6728} {\bibfield  {journal} {\bibinfo
  {journal} {Phys. Rev. B}\ }\textbf {\bibinfo {volume} {47}},\ \bibinfo
  {pages} {6728} (\bibinfo {year} {1993})}\BibitemShut {NoStop}%
\bibitem [{\citenamefont {Geiger}\ and\ \citenamefont
  {Smidt}(2022)}]{e3nn_paper}%
  \BibitemOpen
  \bibfield  {author} {\bibinfo {author} {\bibfnamefont {M.}~\bibnamefont
  {Geiger}}\ and\ \bibinfo {author} {\bibfnamefont {T.}~\bibnamefont {Smidt}},\
  }\bibfield  {title} {\bibinfo {title} {e3nn: Euclidean neural networks},\
  }\bibfield  {journal} {\bibinfo  {journal} {arXiv preprint arXiv:2207.09453}\
  }\href {https://doi.org/10.48550/arXiv.2207.09453}
  {10.48550/arXiv.2207.09453} (\bibinfo {year} {2022})\BibitemShut {NoStop}%
\bibitem [{\citenamefont {Alappat}\ \emph {et~al.}(2020)\citenamefont
  {Alappat}, \citenamefont {Basermann}, \citenamefont {Bishop}, \citenamefont
  {Fehske}, \citenamefont {Hager}, \citenamefont {Schenk}, \citenamefont
  {Thies},\ and\ \citenamefont {Wellein}}]{pardiso-8.0a}%
  \BibitemOpen
  \bibfield  {author} {\bibinfo {author} {\bibfnamefont {C.}~\bibnamefont
  {Alappat}}, \bibinfo {author} {\bibfnamefont {A.}~\bibnamefont {Basermann}},
  \bibinfo {author} {\bibfnamefont {A.~R.}\ \bibnamefont {Bishop}}, \bibinfo
  {author} {\bibfnamefont {H.}~\bibnamefont {Fehske}}, \bibinfo {author}
  {\bibfnamefont {G.}~\bibnamefont {Hager}}, \bibinfo {author} {\bibfnamefont
  {O.}~\bibnamefont {Schenk}}, \bibinfo {author} {\bibfnamefont
  {J.}~\bibnamefont {Thies}},\ and\ \bibinfo {author} {\bibfnamefont
  {G.}~\bibnamefont {Wellein}},\ }\bibfield  {title} {\bibinfo {title} {A
  recursive algebraic coloring technique for hardware-efficient symmetric
  sparse matrix-vector multiplication},\ }\href
  {https://doi.org/10.1145/3399732} {\bibfield  {journal} {\bibinfo  {journal}
  {ACM Trans. Parallel Comput.}\ }\textbf {\bibinfo {volume} {7}},\ \bibinfo
  {pages} {37} (\bibinfo {year} {2020})}\BibitemShut {NoStop}%
\bibitem [{\citenamefont {Bollh{\"o}fer}\ \emph {et~al.}(2019)\citenamefont
  {Bollh{\"o}fer}, \citenamefont {Eftekhari}, \citenamefont {Scheidegger},\
  and\ \citenamefont {Schenk}}]{pardiso-8.0c}%
  \BibitemOpen
  \bibfield  {author} {\bibinfo {author} {\bibfnamefont {M.}~\bibnamefont
  {Bollh{\"o}fer}}, \bibinfo {author} {\bibfnamefont {A.}~\bibnamefont
  {Eftekhari}}, \bibinfo {author} {\bibfnamefont {S.}~\bibnamefont
  {Scheidegger}},\ and\ \bibinfo {author} {\bibfnamefont {O.}~\bibnamefont
  {Schenk}},\ }\bibfield  {title} {\bibinfo {title} {Large-scale sparse inverse
  covariance matrix estimation},\ }\href {https://doi.org/10.1137/17M1147615}
  {\bibfield  {journal} {\bibinfo  {journal} {SIAM Journal on Scientific
  Computing}\ }\textbf {\bibinfo {volume} {41}},\ \bibinfo {pages} {A380}
  (\bibinfo {year} {2019})}\BibitemShut {NoStop}%
\end{thebibliography}%

\end{document}